\documentclass[11pt]{article}

\usepackage{graphicx,latexsym,amssymb,amsmath}

\usepackage{cite} 

\newcommand{\Fig}[1]{Fig.\ref{#1}}

\newcommand{\edo}{\end{document}}

\newcommand{\fractext}[2]{(#1)/(#2)}

\newcommand{\pic}[2]{\includegraphics[scale=#1]{#2}}

\newcommand{\R}{{\mathbb R}}  

\newtheorem{itlemma}{Lemma}[section] 

\newtheorem{itremark}[itlemma]{Remark}

\newenvironment{myremark}{\begin{itremark}\rm}{\end{itremark}} 

\newcommand{\be}[1]{\begin{equation}\label{#1}}
\newcommand{\ee}{\end{equation}}
\newcommand{\bl}[1]{\begin{lemma}\label{#1}}
\newcommand{\ble}[1]{\begin{lemmaex}\label{#1}}
\newcommand{\br}[1]{\begin{myremark}\label{#1}}
\newcommand{\bt}[1]{\begin{theorem}\label{#1}}
\newcommand{\bd}[1]{\begin{definition}\label{#1}}
\newcommand{\bp}[1]{\begin{proposition}\label{#1}}
\newcommand{\bc}[1]{\begin{corollary}\label{#1}}
\newcommand{\bfact}[1]{\begin{fact}\label{#1}}
\newcommand{\ber}[1]{\begin{exercise}\label{#1}}
\newcommand{\bex}[1]{\begin{example}\label{#1}}
\newcommand{\bem}[1]{\begin{example}\label{#1}}  
\newcommand{\ec}{\mybox\end{corollary}}
\newcommand{\efact}{\mybox\end{fact}}
\newcommand{\eer}{\mybox\end{exercise}}
\newcommand{\eex}{\mybox\end{example}}
\newcommand{\eem}{\mybox\end{example}}
\newcommand{\el}{\mybox\end{lemma}}
\newcommand{\ele}{\mybox\end{lemmaex}}
\newcommand{\er}{\mybox\end{myremark}}
\newcommand{\et}{\qed\end{theorem}}
\newcommand{\ed}{\mybox\end{definition}}
\newcommand{\ep}{\mybox\end{proposition}}
\newcommand{\epr}{\end{proof}}
\newcommand{\bpr}{\begin{proof}}

\newcommand{\ecs}{\end{corollary}}
\newcommand{\eers}{\end{exercise}}
\newcommand{\eexs}{\end{example}}
\newcommand{\eems}{\end{example}}
\newcommand{\els}{\end{lemma}}
\newcommand{\eles}{\end{lemmaex}}
\newcommand{\ers}{\end{remark}}
\newcommand{\ets}{\end{theorem}}
\newcommand{\eds}{\end{definition}}
\newcommand{\eps}{\end{proposition}}

\newcommand{\mybox}{\hfill $\Box$} 
\newcommand{\beq}{\begin{eqnarray}}
\newcommand{\eeq}{\end{eqnarray}}
\newcommand{\beqn}{\begin{eqnarray*}}
\newcommand{\eeqn}{\end{eqnarray*}}
\newcommand{\bi}{\begin{itemize}}
\newcommand{\ei}{\end{itemize}}
\newcommand{\ben}{\begin{enumerate}}
\newcommand{\een}{\end{enumerate}}

\newcommand{\precc}{\prec\!\prec}

\newcommand{\inputu}{u}

\newcommand{\ztot}{z_{\mbox{\tiny tot}}}
\newcommand{\ytot}{y_{\mbox{\tiny tot}}}

\newcommand{\coefV}{v}

\newcommand{\coefK}{k}
\newcommand{\coefk}{v}
\newcommand{\gain}{g}
\newcommand{\minp}[2]{\begin{minipage}{#1\textwidth}#2\end{minipage}}

\begin{document}

\title{Monotone and near-monotone network structure (part II)}
\author{Eduardo D.\ Sontag%
\footnote{Supported in part by NSf Grants DMS-0504557 and DMS-0614371.}\\
Rutgers University, New Brunswick, NJ, USA\\
\texttt{sontag@math.rutgers.edu}}
\maketitle

\medskip
(continued from Part I)
\medskip

\protect\addtocounter{section}{2}
\section{I/O Monotone systems}
\label{mios-section}

We next describe recent work on \emph{monotone input/output systems}
(``\emph{MIOS}'' from now on).  Monotone i/o systems originated in the
analysis of mitogen-activated protein kinase cascades and other cell signaling
networks, but later proved useful in the study of a broad variety of other
biological models.
Their surprising breath of applicability notwithstanding, of course MIOS
constitute a restricted class of models, especially when seen in the context
of large biochemical networks.  Indeed, the original motivation for
introducing MIOS, in the 2003 paper~\cite{monotoneTAC}, was to study an
existing \emph{non-monotone} model of negative feedback in MAPK cascades.
The key breakthrough was the realization that this example, and, as it turned
out, many others, can be profitably studied by \emph{decompositions into
  MIOS}.  In other words, a non-monotone system is viewed as an
interconnection of monotone subsystems.  Based on the architecture of the
interconnections between the subsystems (``network structure''), one deduces
properties of the original, non-monotone, system.  (Later work, starting
with~\cite{monotoneMulti}, showed that even monotone systems can be usefully
studied through this decomposition-based approach.)

We review the basic notion from~\cite{monotoneTAC}.
(For concreteness, we make definitions for systems of ordinary differential
equations, but similar definitions can be given for abstract dynamical systems,
including in particular reaction-diffusion partial differential equations and
delay-differential systems, see e.g.~\cite{dcds06}.)
The basic setup is that of an input/output system in the sense of mathematical
systems and control theory \cite{mct}, that is, sets of equations
\be{iosys}
\frac{dx}{dt}=f(x,u),\;\; y=h(x) \,,
\ee
in which states $x(t)$ evolve on some subset $X\subseteq \R^n$, and input and output
values $u(t)$ and $y(t)$ belong to subsets $U\subseteq \R^m$ and $Y\subseteq \R^p$ 
respectively. 
The coordinates $x_1,\ldots ,x_n$ of states typically represent
concentrations of chemical species, such as proteins, mRNA, or metabolites.
The input variables, which can be seen as controls, forcing functions, or
external signals, act as stimuli. 
Output variables can be thought of as describing
responses, such as movement, or
as measurements provided by biological reporter devices like GFP that allow a
partial read-out of the system state vector $(x_1,\ldots ,x_n)$.
The maps $f:X\times U\rightarrow \R^n$ and $h:X\rightarrow Y$ are taken to be continuously
differentiable.  (Much less can be assumed for many results, so long as local
existence and uniqueness of solutions is guaranteed.)
An \emph{input} is a signal $u:[0,\infty )\rightarrow U$ which is locally essentially compact
(meaning that images of restrictions to finite intervals are compact), and
we write $\varphi(t,x_0,u)$ for the solution of the initial value problem
$dx/dt(t)=f(x(t),u(t))$ with $x(0)=x_0$, or just $x(t)$ if $x_0$ and $u$ are
clear from the context, and $y(t)=h(x(t))$.
See~\cite{mct} for more on i/o systems.
For simplicity of exposition, we make the blanket assumption that solutions do
not blow-up on finite time, so $x(t)$ (and $y(t)$) are defined for all $t\geq 0$.
(In biological problems, almost always conservation laws and/or boundedness
of vector fields insure this property.  In any event, extensions to local
semiflows are possible as well.)

Given three partial orders on $X,U,Y$
(we use the same symbol $\prec$ for all three orders),
a monotone I/O system (MIOS), with
respect to these partial orders,
is a system~(\ref{iosys}) such that $h$ is a
monotone map (it preserves order) and:
for all initial states $x_1,x_2$
for all inputs $u_1,u_2$,
the following property holds:
if $x_1$$\preceq$$x_2$
and $u_1$$\preceq$$u_2$
(meaning that 
$u_1(t)$$\preceq$$u_2(t)$ for all $t$$\geq $$0$), then
$\varphi(t,x_1,u)$$\preceq$$\varphi(t,x_2,u_2)$
for all $t>0$.
Here we consider partial orders induced by closed proper cones
$K\subseteq \R^\ell$, in the sense that $x\preceq y$ iff $y-x\in K$.
The cones $K$ are assumed to have a nonempty interior and are pointed, i.e.\
$K\bigcap -K=\{0\}$.
A \emph{strongly} monotone system is one which satisfies the following
stronger property:
if $x_1\preceq x_2$ and $u_1\preceq u_2$, then the strict inequality
$\varphi(t,x_1,u)\precc \varphi(t,x_2,u_2)$ holds for all $t>0$, where
$x\precc y$ means that $y-x$ is in the interior of the cone $K$.

The most interesting particular case is that in which $K$ is 
an \emph{orthant} cone in $\R^n$, i.e.\ a set $S_\varepsilon $ of the form 
$\{x\in \R^n\,|\, \varepsilon _i x_i\geq 0\}$, where $\varepsilon _i=\pm 1$ for each $i$.

When there are no inputs nor outputs, the definition of monotone systems
reduces to the classical one of monotone dynamical systems studied by Hirsch,
Smith, and others \cite{smith}.  This is what we discussed
earlier, for the case of orthant cones.
When there are no inputs, strongly monotone classical systems have especially
nice dynamics.  Not only is chaotic or other irregular behavior ruled out,
but, in fact, almost all bounded
trajectories converge to the set of steady states
(Hirsch's generic convergence theorem~\cite{Hirsch,Hirsch2}).

A useful test for monotonicity with respect to orthant cones,
which generalizes Kamke's condition to the i/o case, is
as follows.
Let us assume that all the
partial derivatives $\frac{\partial f_i}{\partial x_j}(x,u)$ for
$i\not= j$, $\frac{\partial f_i}{\partial u_j}(x,u)$ for all $i,j$, and 
$\frac{\partial h_i}{\partial x_j}(x)$ for all $i,j$
(subscripts indicate components)
do not change sign, i.e., they are either always $\geq 0$ or always $\leq 0$.
We also assume that $X$ is convex (much less is needed.)
We then associate a directed graph $G$ to the given MIOS, 
with $n+m+p$ nodes, and edges labeled ``$+$'' or ``$-$'' (or $\pm1$),
whose labels are determined by the signs of the appropriate partial
derivatives (ignoring diagonal elements of $\partial f/\partial x$).
One may define in an obvious manner undirected loops in $G$, and
the \emph{parity} of a loop is defined
by multiplication of signs along the loop.
(See e.g.~\cite{monotoneMulti,monotoneLSU} for more details.)
Then, it is easy to show that a system is monotone with respect to \emph{some}
orthant cones in $X,U,Y$ if and only if there are no negative loops in $G$.
A sufficient condition for strong monotonicity is that, in addition to
monotonicity, the partial Jacobians of $f$ with respect to $x$ should be
everywhere irreducible .
(``Almost-everywhere'' often suffices; see
\cite{smith,Hirsch-Smith}.  See these references also for
extensions to non-orthant cones in the case of no inputs and outputs, based on
work of Schneider and Vidyasagar, Volkmann, 
and others~\cite{Schneider_Vidyasagar,Volkman,Walcher,Walter}).

In inhibitory feedback, a chemical species $x_j$ typically affects the
rate of formation of another species $x_i$ through a term like
$h(x_j)={V}/({K+x_j})$.  The decreasing function $h(x_j)$ can be seen as the
output of an \emph{anti-monotone} system, i.e. a system which satisfies the
conditions for monotonicity, except that the output
map \emph{reverses} order: $x_1\preceq x_2 \Rightarrow h(x_2)\preceq h(x_1)$.

An interconnection of monotone subsystems, that is to say, an entire
system made up of monotone components, may or may not be monotone: ``positive
feedback'' (in a sense that can be made precise) preserves monotonicity,
while ``negative feedback'' destroys it.
Thus, oscillators such as circadian rhythm generators require negative
feedback loops in order for periodic orbits to arise, and hence are
not themselves monotone systems, although they can be decomposed into
monotone subsystems (cf.~\cite{04cdc-circadian}).
A rich theory is beginning to
arise, characterizing the behavior of non-monotone interconnections.
For example, \cite{monotoneTAC} shows how to preserve convergence to
equilibria; see also the follow-up papers
\cite{angelileenheersontagSCL04,predatorpreysgt05,dcds06,enciso_smith_sontagJDE06,gedeon05}.
Even for monotone interconnections, the decomposition approach is very
useful, as it permits locating and characterizing the stability of
steady states based upon input/output behaviors of components,
as described in~\cite{monotoneMulti};
see also the follow-up papers
\cite{pnasangeliferrellsontag04,enciso_sontagSCL05,leenheer-malisoff}.

Moreover, a key point brought up in~\cite{monotoneTAC,monotoneMulti,sysbio04,ejc05es} is that new techniques for
monotone systems in many situations allow one to characterize the behavior of
an entire system, based upon the ``qualitative'' knowledge represented by
general network topology and the inhibitory or activating character of
interconnections, combined with only a relatively \emph{small amount of
quantitative} data.  The latter data may consist of steady-state responses of
components (dose-response curves and so forth), and there is no need to know
the precise form of dynamics or parameters such as kinetic constants in order
to obtain global stability conclusions and study global bifurcation behavior.
We now discuss these issues.

\subsubsection*{Characteristics}

The main results in~\cite{monotoneTAC,monotoneMulti} were built around the
study of \emph{characteristics}, also called \emph{step-input steady-state
responses} or (nonlinear) \emph{DC gains}.
To explain this concept, we study the effect of a \emph{constant} input
$u(t)\equiv u_0, t\geq 0$ (in biological terms, the constant input may represent
the extracellular concentration of a ligand during a particular experiment,
for example).
For each such constant input, we study the
dynamical system $dx/dt=f(x,u_0)$.
Let us assume that all the solutions of this system approach steady states,
and let us call $K(u_0)$ the set of steady states that arises in this way.
To each state $x$ in this set $K(u_0)$, one may associate the corresponding
output or measured quantity $h(x_0)$.  Let $k(u_0)$ be the set of all output
values that arise in this manner.
The graph of the set-valued mapping $u_0\mapsto k(u_0)$ is a subset of the cross
product space $\R^m\times \R^p$, which may be though of as a curve when $m=p=1$,
and which describes the possible steady state output values for any given
constant input.

Although many results may be given in more generality, we will assume for the
remainder of this paper that
these mappings are single-valued, not set-valued, in other words that
the system is monostable.  Thus,
a (single-valued) characteristic is said to exist for the system if there is 
a unique steady state for the dynamical system $dx/dt=f(x,u_0)$, denoted
$K(u_0)$, and this property is true for all possible constant levels $u_0$.
We then define the (output) \emph{characteristic} $k:U\rightarrow Y$ as the composition
$h\circ K$.
Under reasonable assumptions on $X$ and boundedness, appealing to results
from~\cite{JiFa,dancer98} allows one to conclude that $K(u_0)$ is in fact
a globally asymptotically stable (``GAS'' from now on) state for
$dx/dt=f(x,u_0)$, so that all trajectories (for this ``frozen'' value of the
input $u$), converge to $K(u_0)$, and the output $y(t)$ converges to $k(u_0)$.

Characteristics (dose response curves, activity plots, steady-state expression
of a gene in response to an external ligand, etc.) are frequently available
from experimental data, especially in 
molecular biology and pharmacology (for instance, in
the modeling of receptor-ligand interactions~\cite{receptorligandJTB04}).
A goal of MIOS analysis is to \emph{combine the numerical
information provided by characteristics with the qualitative information given
by (signed) network topology in order to predict global bifurcation
behavior.}
(See~\cite{ejc05es} for a longer discussion of this
``qualitative-quantitative approach'' to systems biology.)
On the other hand, characteristics are also a very powerful tool for the purely
mathematical analysis of existing models, as we show below.
Monotone systems with well-defined characteristics constitute a
very well-behaved set of building blocks for arbitrary systems.
In particular, cascades 
of such systems inherit the same properties (monotone, monostable response).
The original theorems, in the works~\cite{monotoneTAC,monotoneMulti},
dealt with systems obtained by
interconnecting monotone (or anti-monotone) I/O systems with characteristics
in \emph{feedback}.
Let us review them next.

\subsubsection*{Positive feedback}

The first basic theorem refers to a feedback interconnection of two MIOS
\beq
\label{sys1}
\frac{x_1}{dt}&=&f_1(x_1,u_1),\;\; y_1=h_1(x_1)\\
\label{sys2}
\frac{x_2}{dt}&=&f_2(x_2,u_2), \;\; y_2=h_2(x_2)
\eeq
with characteristics denoted by ``$k$'' and ``$g$'' respectively.
(A degenerate case, in which the second system is memory-free, that is,
there are no state variables $x_2$ and $y_2$ is simply a static
function $y_2(t)=g(u_2(t))$, is also allowed.  In fact, the proof of
the general case can be reduced to that of the degenerate case, simply
by taking the first system as a cascade connection of the two systems.)

\begin{figure}[ht]
\centering
\pic{0.12}{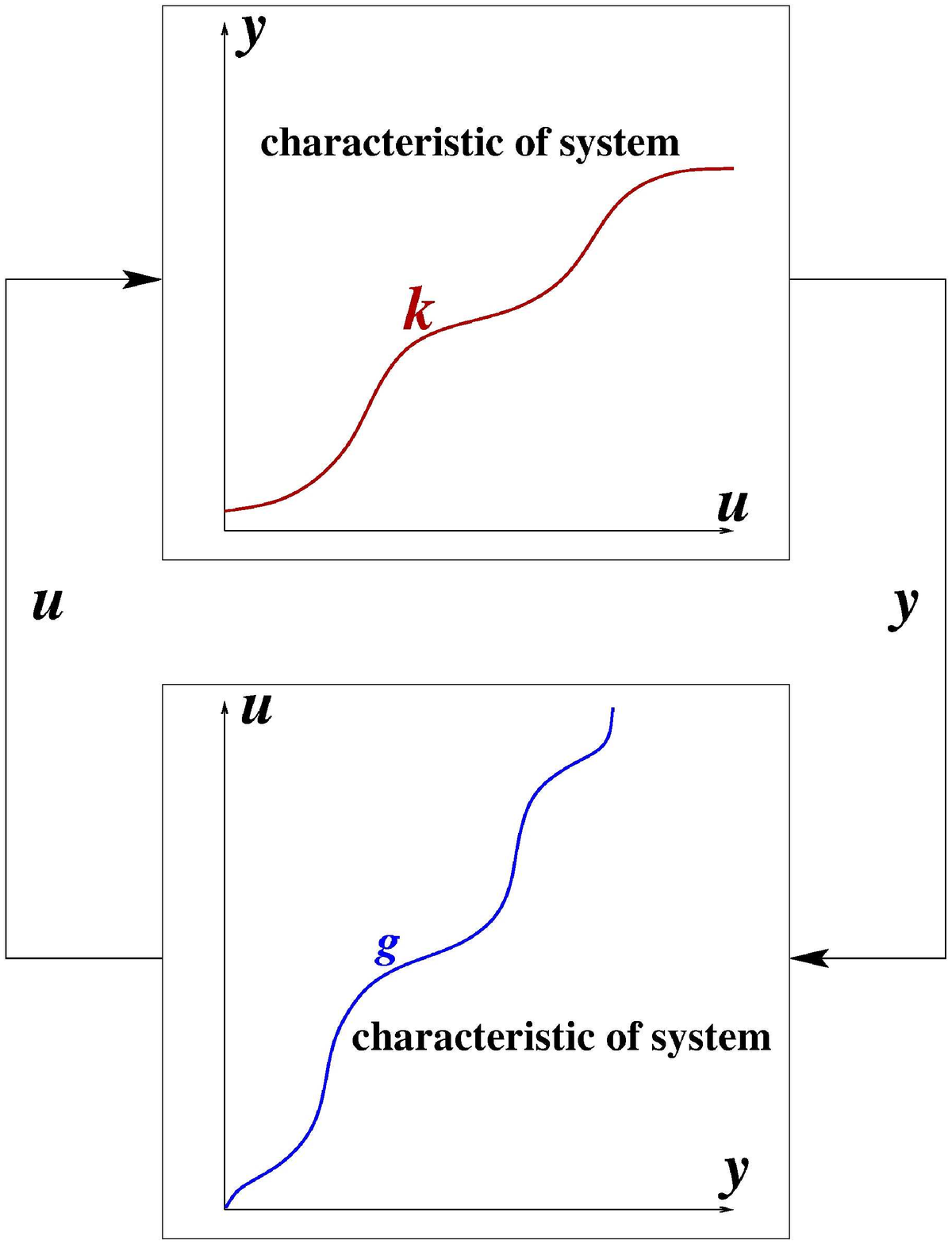}
\pic{0.2}{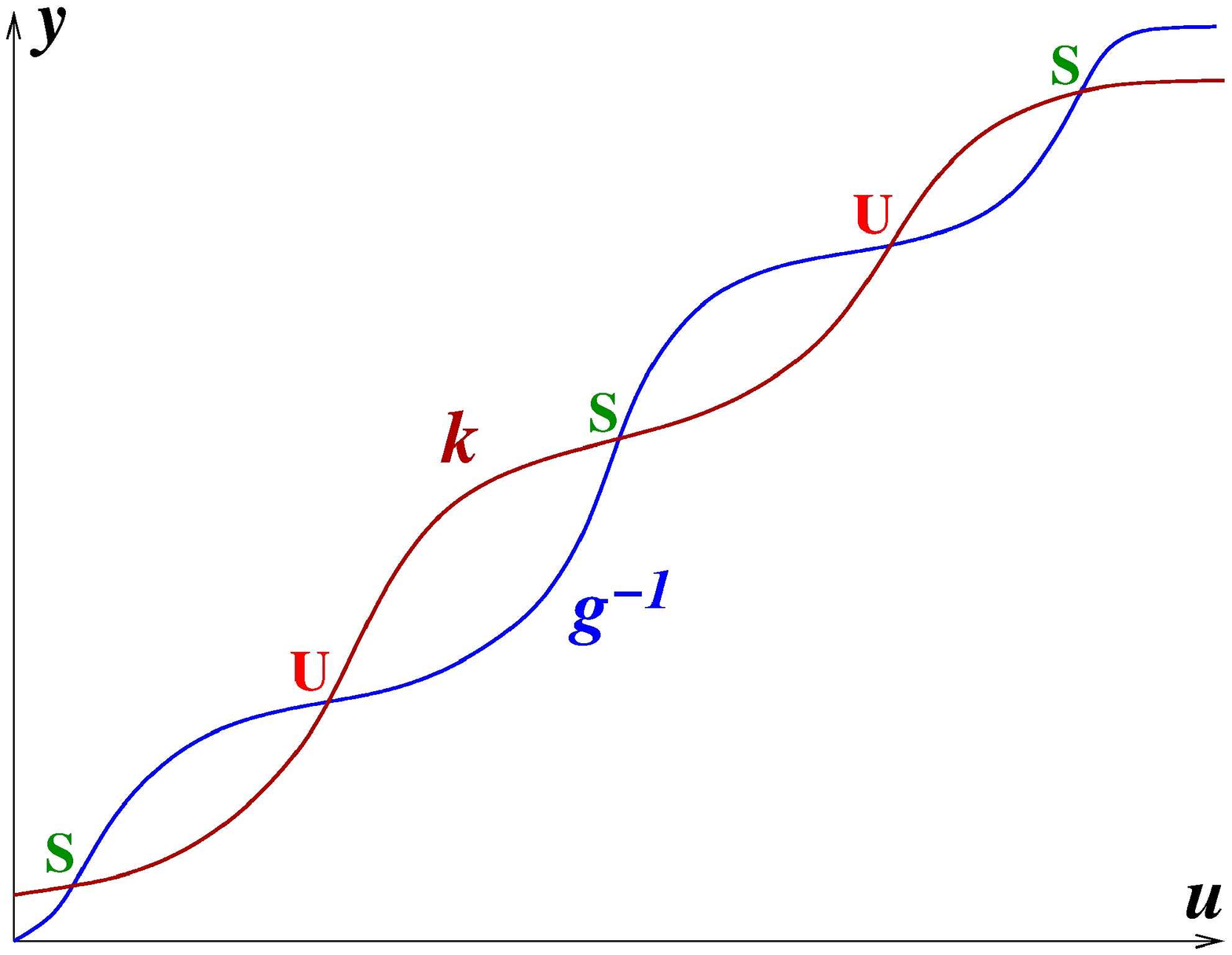}
\pic{0.12}{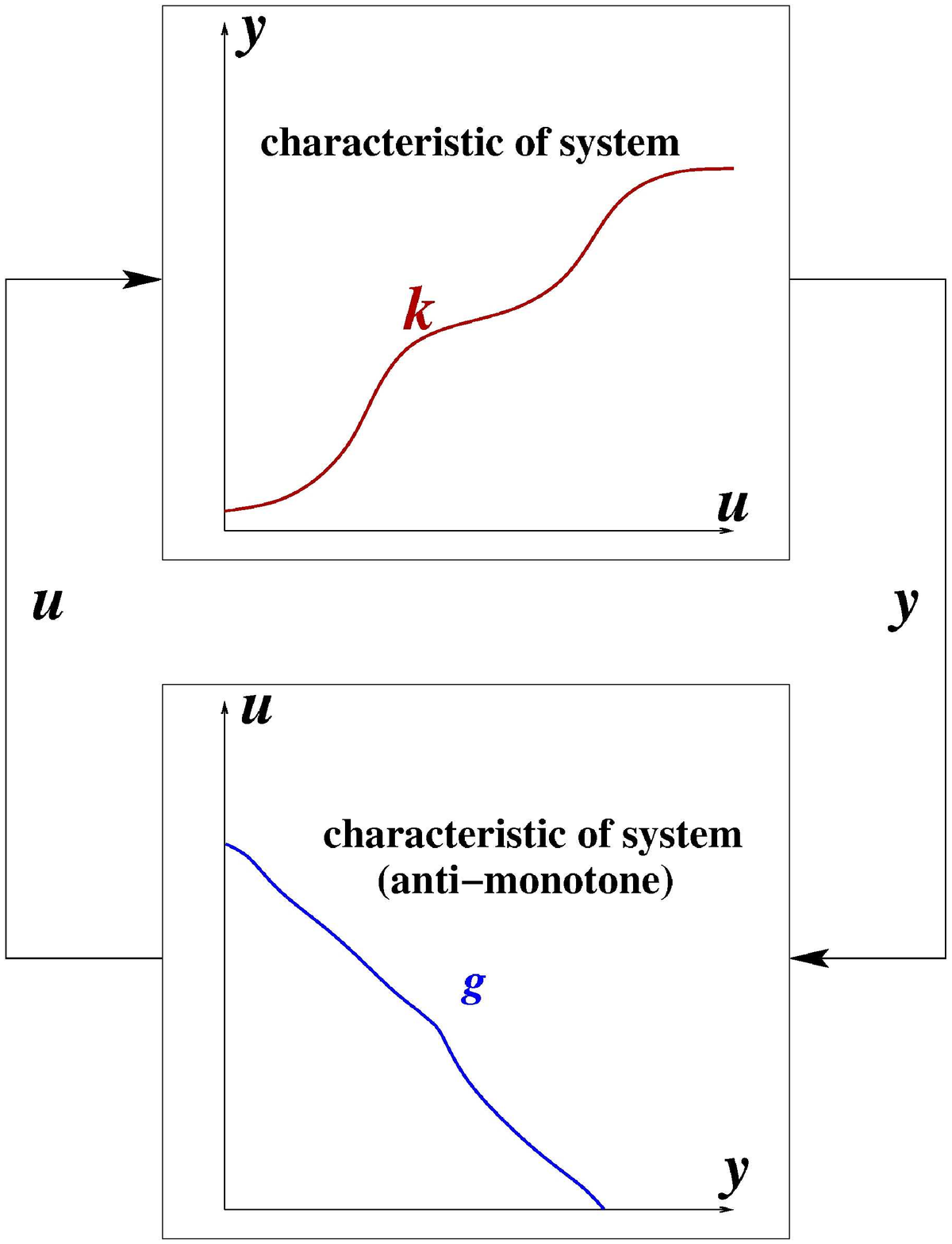}
\pic{0.2}{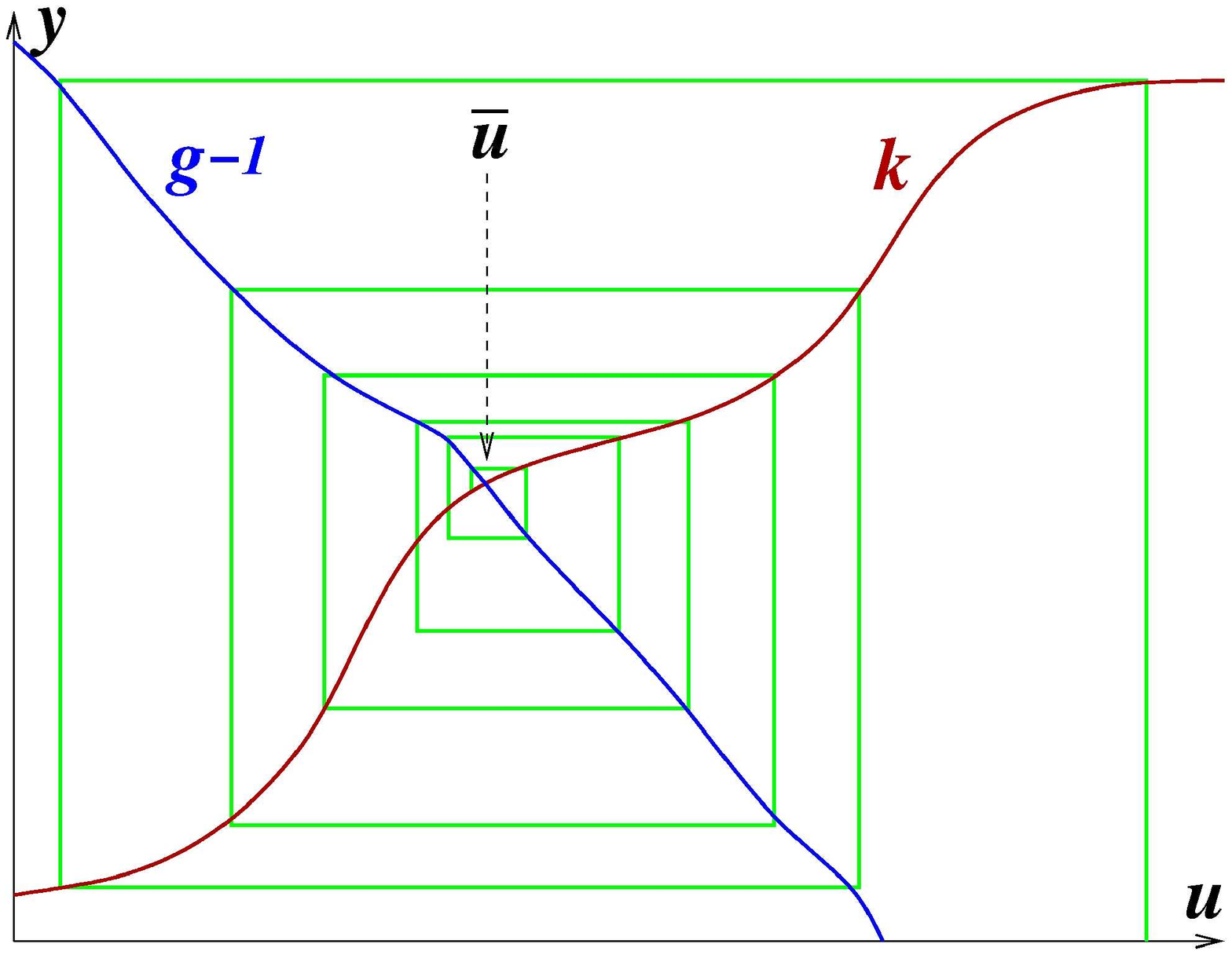}
\caption{(a) Positive feedback
and (b) characteristics;
(c) negative feedback
and
(d) characteristics$\quad\quad$}
\label{oldfigs78}
\end{figure}

As in~\cite{monotoneMulti}, we suppose that the inputs and outputs of both
systems are scalar: $m_1$$=$$m_2$$=$$p_1$$=$$p_2$$=1$ (see~\cite{enciso_sontagSCL05} for a
generalization to high-dimensional inputs and outputs).
The ``positive feedback interconnection'' of the systems~(\ref{sys1})
and~(\ref{sys2}) is defined by letting the output of each of them serve as the
input of the other ($u_2$$=$$y_1$$=$``$y$'' and $u_1$$=$$y_2$$=$``$u$''),
as depicted in Figure~\ref{oldfigs78}(a).
Such positive feedback systems may easily be multi-stable, even if the
constituent pieces are monostable
\cite{thomas81,snoussi,cinquin02,tyson-sniffers}. 
Let us first discuss how multi-stability may arise in a very intuitive and
simple example, and later present the general theorem.

\begin{figure}[h,t]
 \begin{center}
 \pic{0.12}{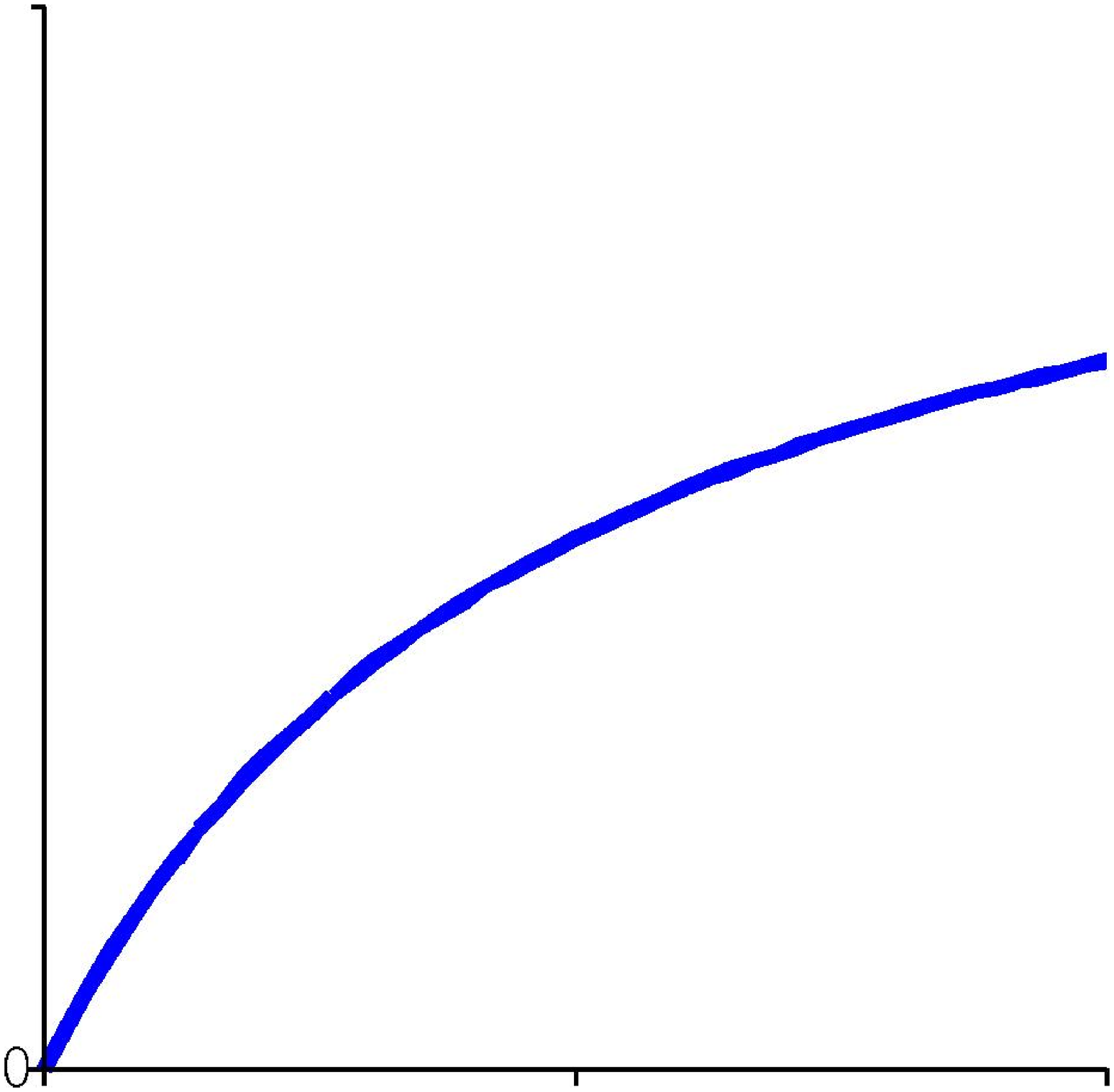} \  \ 
 \pic{0.1}{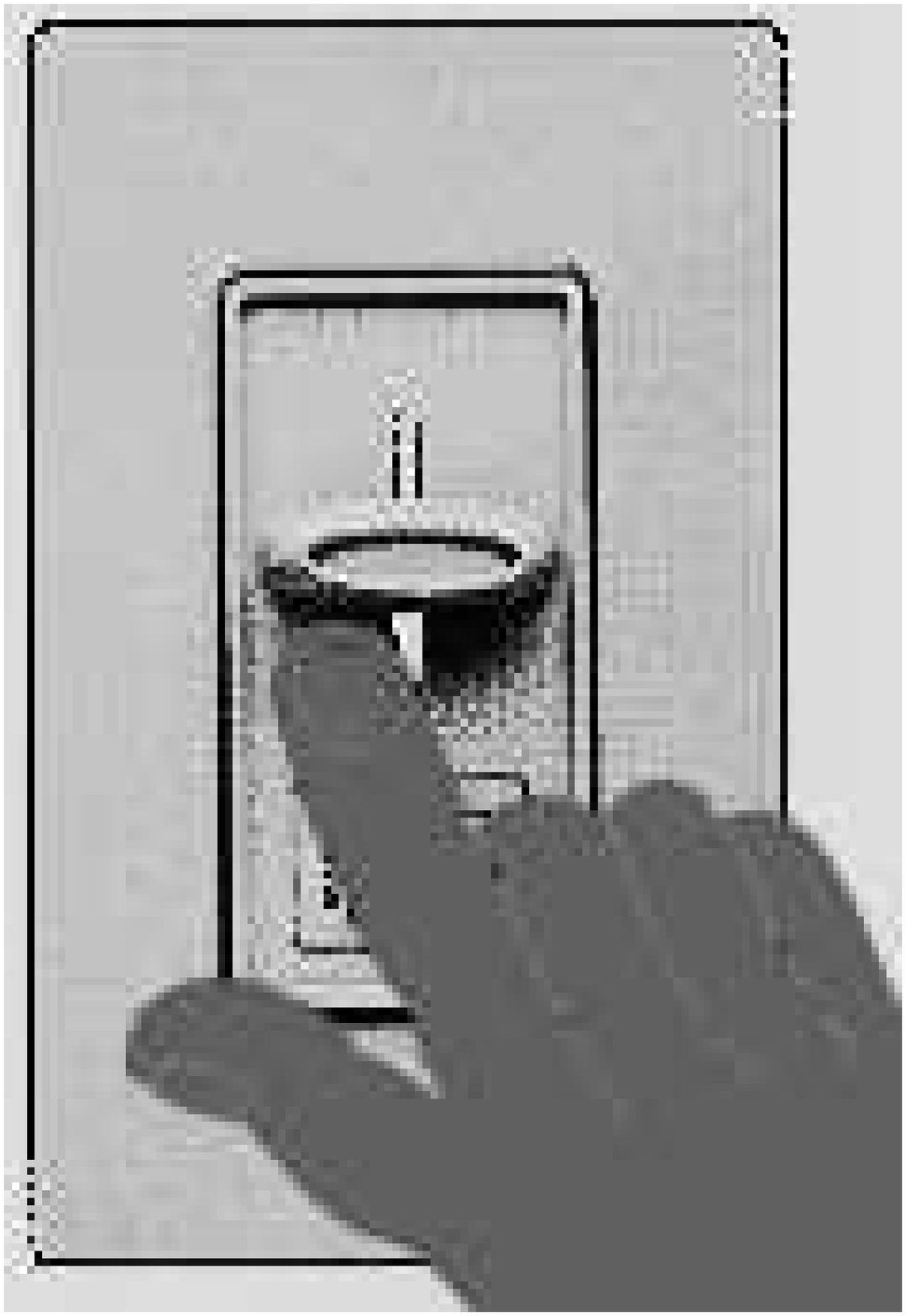} \  \  \ 
 \pic{0.12}{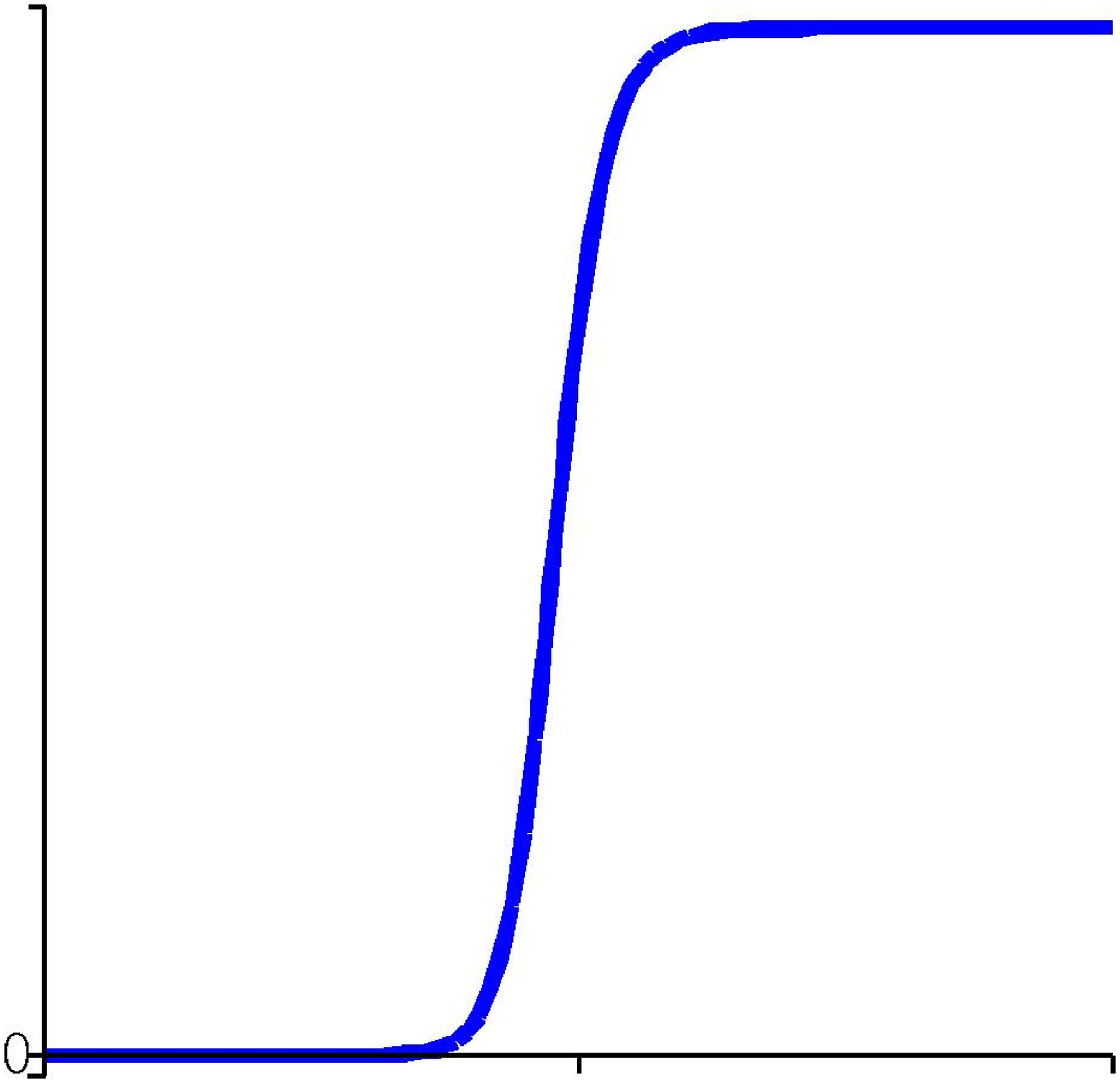} \  \  \ 
 \pic{0.12}{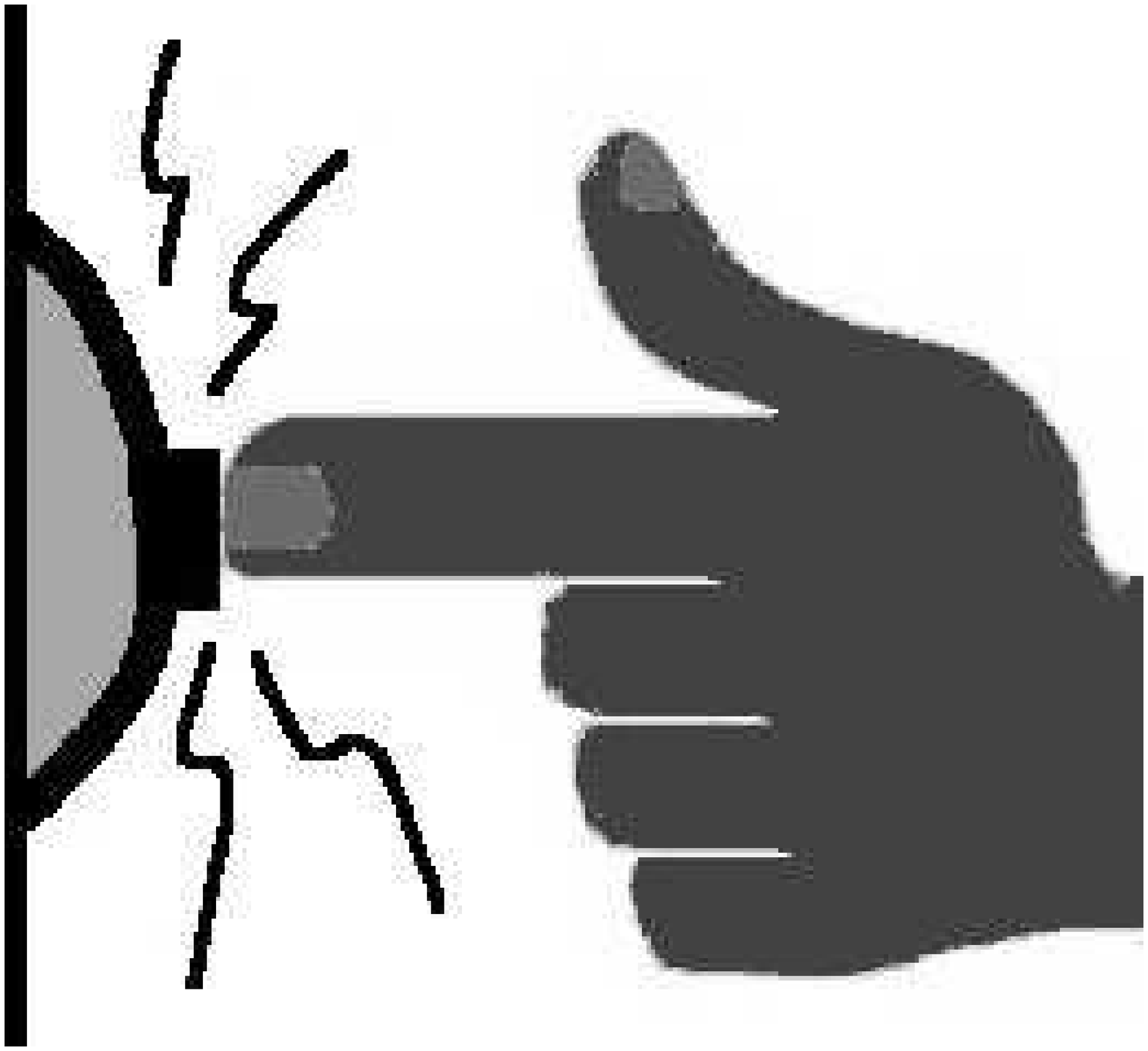} \  \  \ 
 \pic{0.12}{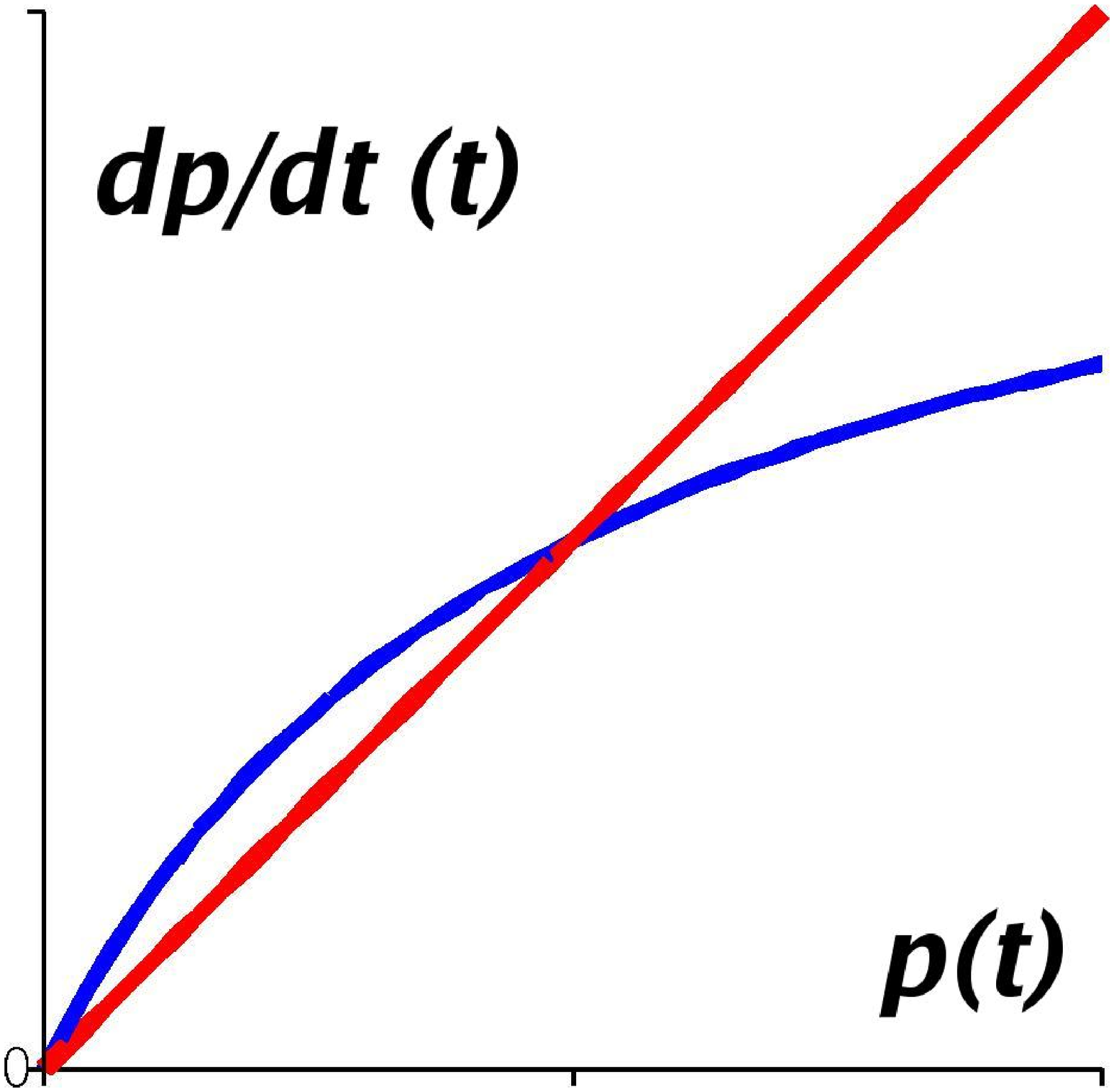} \  \  \ 
 \pic{0.12}{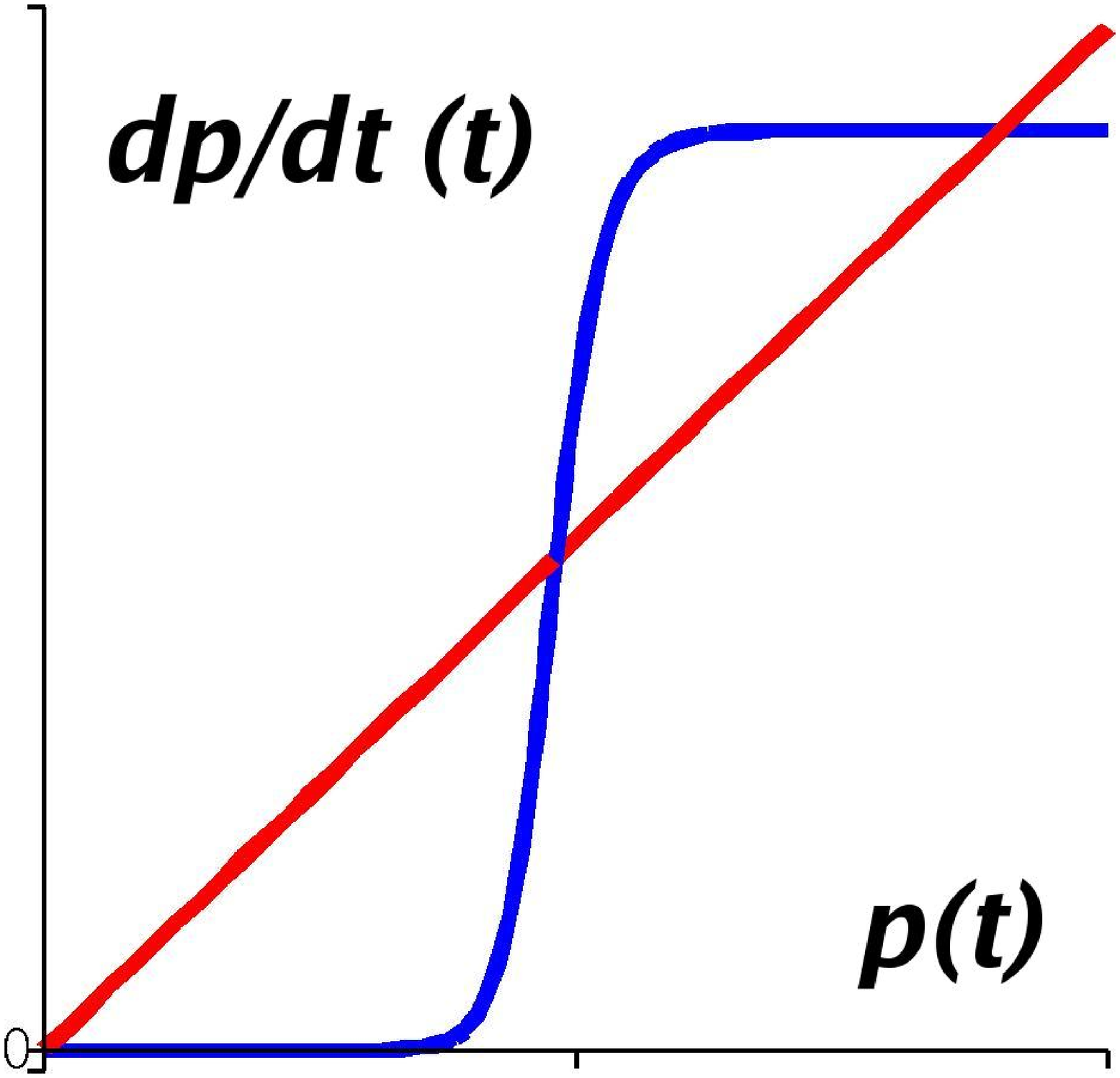}
 \caption{(a) hyperbolic and
 (b) sigmoidal responses; \ \ \ 
 (c) intersections with degradation;
 }%
 \label{hyperbolic_response}
 \end{center}
 \end{figure}

Two typical steady-state responses are as follows.
Suppose that P is a protein with
Michaelis-Menten production rate and linear degradation:
$dp/dt = V_{max}u/(k_m+u) - kp$,
where $u$ represents the concentration a substrate that is used in P's
formation. 
The reporter variable is $y(t)$$=$$p(t)$.
In this case, the steady state when $u(t)$$\equiv $$u_0$ is
$p_0 $$=$$ k(u_0)$$=$$ (V_{max}/k)u_0/(k_m+u_0)$, and we obtain
a {\em hyperbolic\/} response, \Fig{hyperbolic_response}(a).
The response in this example is {\em graded\/} (``light-dimmer''): it is
proportional to the parameter $u_0$ on a large range of values until
saturation.  In contrast,
a {\em sigmoidal\/} (``doorbell'')
response,
\Fig{hyperbolic_response}(b), which may arise from
$dp/dt = V_{max}u^r/(k_m^r+u^r) - kp$
with Hill coefficient (cooperativity index) $r$$>$$1$,
implies that values of $u_0$ under some threshold will not result in an
appreciable activity, while values over the threshold give an abrupt
change (respectively, $p_0\approx 0$ in steady state and
$p_0\approx V_{max}/k$ in steady state).
Sigmoidal responses are critical e.g.\ if a cell must decide in a
binary fashion whether a gene should be transcribed or not, depending on
an extracellular signal
\cite{cit6,thomas01,cit20,novick57,cit5,cit8,cit13,ptashne92,cit14,cit15,cit16,sha03,pomerening-ferrell,cit9,cit13,cit14,cit19}.
Cascades of enzymatic reactions can be made to display such
ultrasensitive
response, if there is a Hill coefficient $r$$>$$1$ at each step
\cite{ferrell-cascades}.

Multiple attractors may appear if the output $y$
(for example $y$$=$$p$ in the example) is fed-back as input $u$.
The mechanism might be an autocatalytic process ($u$$=$$y$, e.g.\ if
$p$ helps promote its own transcription) or via a more complicated positive
feedback pathway from $p$ to $u$.
Formally, substituting $u$$=$$p$ into
$dp/dt=\fractext{V_{max}u^r}{k_m^r+u^r}$$- $$kp$ (where $r$$=$$1$ or
$r$$>$$1$), we obtain the closed-loop equation
$dp/dt=\fractext{V_{max}p^r}{k_m^r+p^r}- kp$.
We plot in \Fig{hyperbolic_response}(c)
both the
first term (formation rate) and the second one (degradation), in cases where
$r$$=$$1$ (left) or $r$$>$$1$ (right).
For $r$$=$$1$, for small $p$ the formation rate is larger than
the degradation rate but for large $p$ the opposite holds,
so the concentration {$p(t)$ converges to a unique intermediate value.}
For $r$$>$$1$,
for small $p$ the degradation rate is larger,
so {$p(t)$ converges to a low value}, but
for large $p$ the formation rate is larger and
{$p(t)$ converges to a high value} instead.
Thus, two stable states are created, one low and one high, by this
interaction of formation and degradation.
(There is also an intermediate, unstable state.)
{\em This reasoning is totally elementary, but it provides an intuition for
the general result in~\cite{monotoneMulti}, shown next, which represents a
far-reaching generalization.}
(The result can also be viewed as generalizing aspects of
the papers~\cite{rapp75,hastings-tyson-webester-periodic-goodwin-jde77,allwright77,othmer76jmb,tyson-othmer,thron,malletparet-smith,gedeon_cyclic,smith,smith-genes},
to arbitrary MIOS.)

We consider Fig.~\ref{oldfigs78}(b), where we have plotted together $k$ and
the inverse of $g$.  
It is quite obvious that there is a bijective correspondence between the
steady states of the feedback system and the intersection points of the two
graphs.
With some mild technical conditions of transversality and ``controllability''
and ``observability''
(the recent papers~~\cite{enciso_multi_submitted,05cdc_enciso_sontag} show that
even these mild conditions can be largely dispensed with), the following
much less obvious facts are true.
We first attach labels to the intersection points between the two graphs
as follows: a label $S$ (respectively, $U$) if the slope of $k$ is smaller
(respectively, larger) than the slope of $g^{-1}$ at the intersection point.
One can then conclude that ``almost all'' (in a measure-theoretic sense or
in a Baire-category sense) bounded solutions of the feedback system must
converge to one of the steady states corresponding to intersection points
labeled with an $S$.
The proof reduces ultimately to an application of Hisrch's generic convergence
theorem to the closed-loop system (the technical conditions insure strong 
monotonicity).  However, the value-added is in the fact that stable states
can be identified merely from the \emph{one-dimensional plot} shown
in Fig.~\ref{oldfigs78}(b).  
(If each subsystem would have dimension just one, one can also interpret the
result in terms of a simple nullcline analysis; see the Supplementary Section
of~\cite{pnasangeliferrellsontag04}.)
We remark that the theorems remain true even if arbitrary delays are allowed
in the feedback loop and/or if space-dependent models are considered and
diffusion is allowed (see~\cite{ejc05es} for a discussion).
A new approach~\cite{angeli-ccw}, based not on monotone theory but on a
notion of ``counterclockwise dynamics,'' extends in a different direction the
range of applicability of this methodology.

We wish to emphasize the potential practical relevance of this result
(and others such as~\cite{angeli-ccw}).  The equations
describing each of the systems are often poorly, or not at all, known.  But, as
long as we can assume that each subsystem is monotone and uni-stable, we can
use the information from the planar plots in Fig.~\ref{oldfigs78}(b) to
completely understand the dynamics of the closed-loop system, no matter how
large the number of state variables.  It is often said that the field of
molecular systems biology is characterized by a {\em data-rich/data-poor\/}
paradox: while on the one hand a huge amount of {\em qualitative\/} network
(schematic modeling) knowledge is available for
signaling, metabolic, and gene regulatory networks, on the other hand little
of this knowledge is {\em quantitative\/}, at least at the level of precision
demanded by most mathematical tools of analysis.
On the other hand, input/output steady state data (from a signal such as a
ligand, to a reporter variable such as the expression of a gene monitored by
GFP, or the activity of a protein measured by a Western blot) is
frequently available.
The problem of exploiting qualitative knowledge, and effectively integrating
relatively sparse quantitative data, is among the most challenging issues
confronting systems biology.  
The MIOS approach provides one way to combine these two types of data.
(For further discussion of this ``data-rich/data-poor'' issue,
see~\cite{sysbio04,ejc05es}.) 

The theorem from~\cite{monotoneMulti} and its generalizations, as well as the
negative feedback result discussed below, amount to a model-reduction
approach.  The bifurcation behavior of the complete closed-loop system is
obtained from a low-order reduction (just to two one-dimensional systems,
connected in feedback, when $m=p=1$) and information on the i/o behavior of
the components.  This model-reduction view is further developed
in~\cite{enciso_sontagSCL05}.

\subsubsection*{More discussion through an example: MAPK cascades}

\begin{figure}[h,t]
 \begin{center}
\pic{0.15}{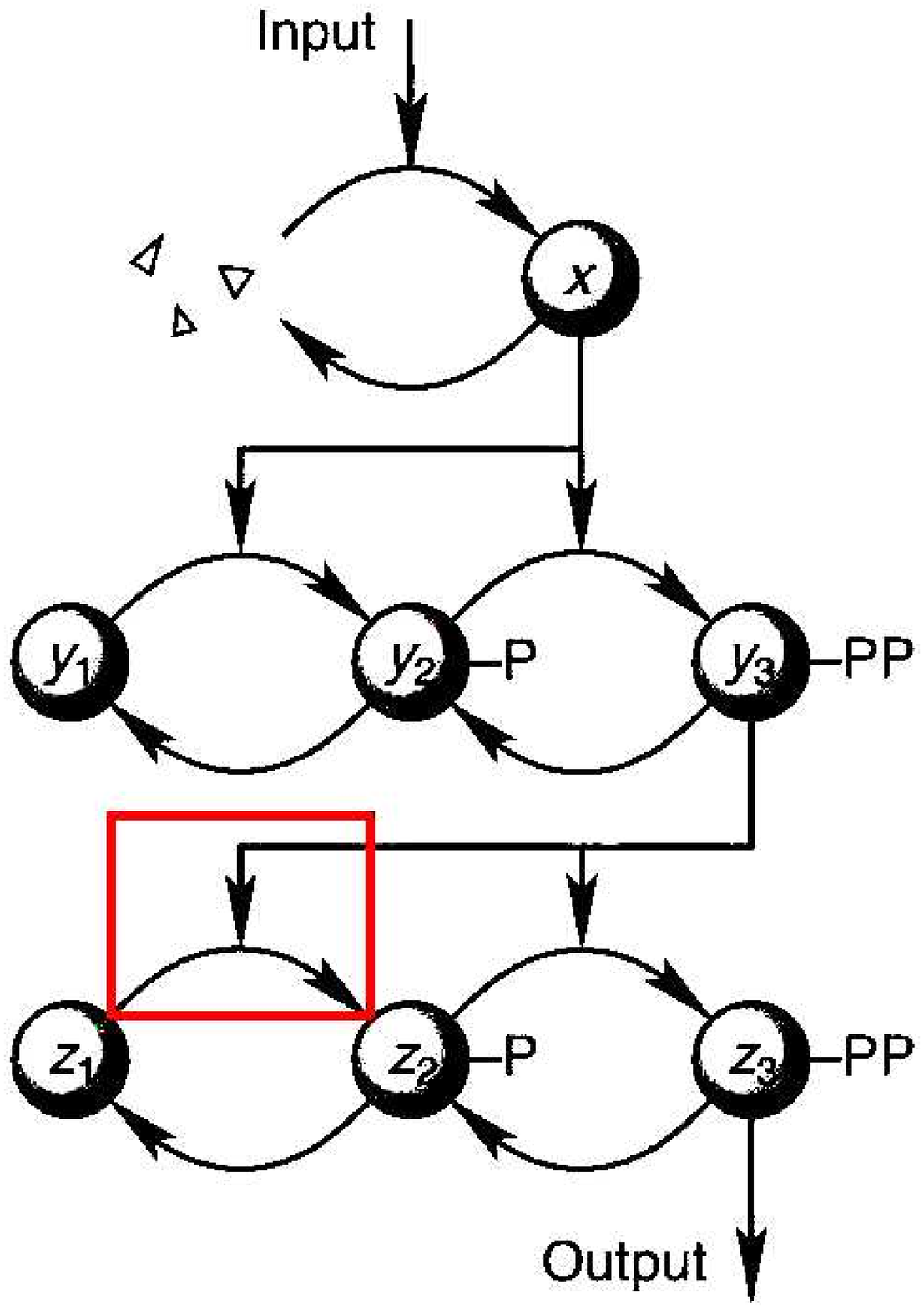}
$\quad\quad$
\setlength{\unitlength}{2000sp}%
\begin{picture}(2416,3924)(3593,-3373)
\thicklines
\put(4801,-361){\circle{600}}
\put(3901,-1261){\circle{600}}
\put(5701,-1261){\circle{600}}
\put(3901,-2461){\circle{600}}
\put(5701,-2461){\circle{600}}
\put(4801,539){\vector( 0,-1){600}}
\put(5701,-2761){\vector( 0,-1){600}}
\put(5701,-1561){\vector( 0,-1){600}}
\put(4201,-1111){\vector( 1, 0){1200}}
\put(5401,-1411){\vector(-1, 0){1200}}
\put(4201,-2311){\vector( 1, 0){1200}}
\put(5401,-2611){\vector(-1, 0){1200}}
\put(4576,-586){\vector(-1,-1){450}}
\put(5026,-586){\vector( 1,-1){450}}
\put(5476,-1561){\vector(-2,-1){1350}}
\put(4500,-3000){output}
\put(4600,-3286){(``$y$'')}
\put(4726,-436){$x$}
\put(3790,-1336){$y_1$}
\put(5590,-1336){$y_3$}
\put(3790,-2536){$z_1$}
\put(5590,-2536){$z_3$}
\put(3900,314){input}
\put(3900,0){(``$u$'')}
{\put(4876,239){$+$}
\put(5776,-3061){$+$}
\put(5251,-736){$+$}
\put(5776,-1861){$+$}
\put(4100,-736){$-$}
\put(4726,-1861){$-$}
\put(4726,-1111){$-$}
\put(4711,-1546){$-$}
\put(4726,-2311){$-$}
\put(4711,-2780){$-$}}
\end{picture} $\quad\quad$
\pic{0.25}{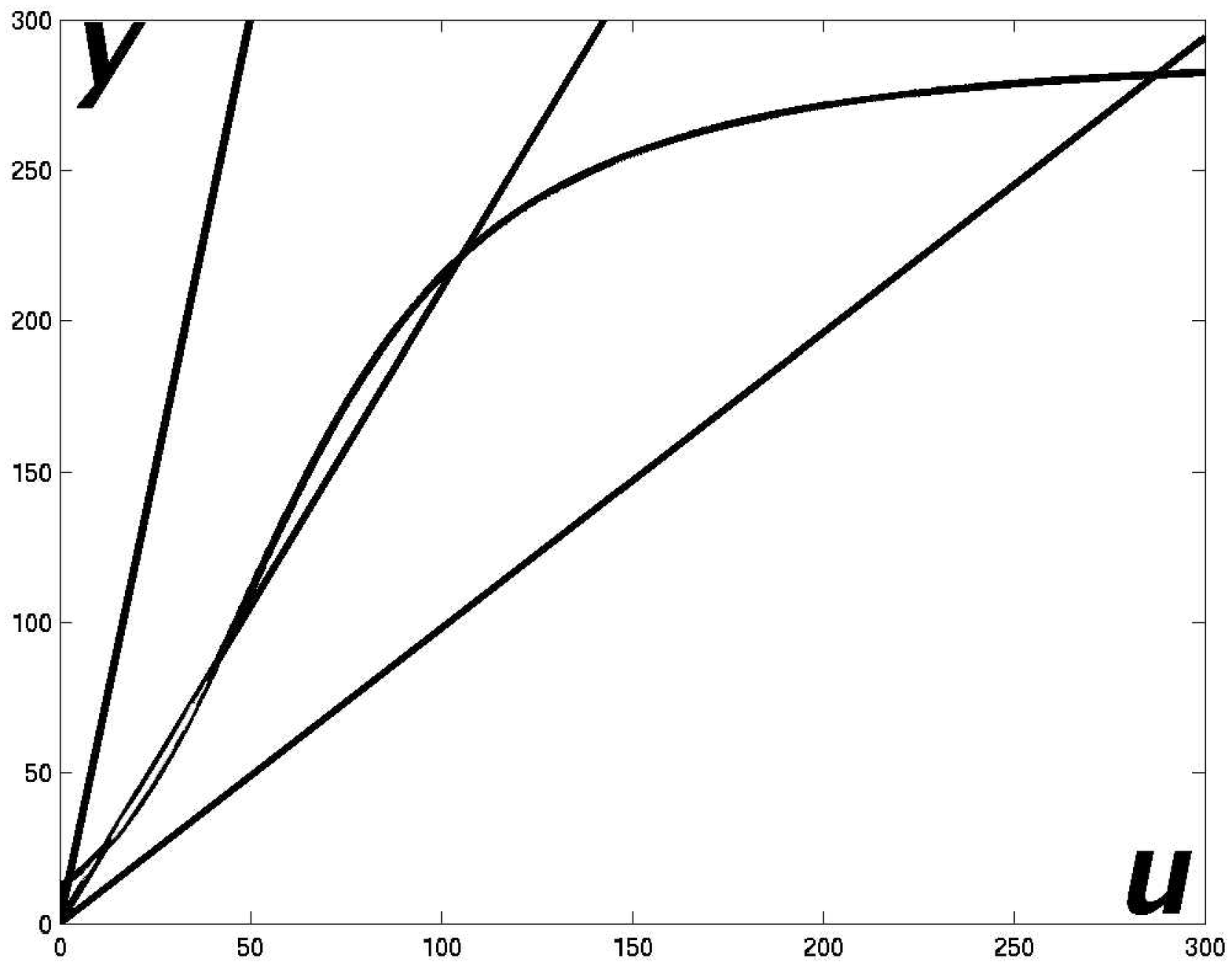}
 \caption{%
  (a) MAPK cascades;
  (b) graph;
  (c) characteristic}
 \label{mapk_cascades_cell_signaling075_cropped}
 \end{center}
 \end{figure}

{\em Mitogen-Activated Protein Kinase (MAPK) cascades\/}
are a ubiquitous ``signaling module'' in eukaryotes, involved in
proliferation, differentiation, development, movement, apoptosis, and
other processes \cite{ferrell,lauffenburger,widman}.
There are several such cascades,
sharing the property of being composed of a cascade
of three kinases.
The basic rule is that two proteins, called generically MAPK and MAPKK
(the last K is for ``kinase of MAPK,'' which is itself a kinase), are
active when doubly phosphorylated, and MAPKK phosphorylates MAPK when
active.  Similarly, a kinase of MAPKK, MAPKKK, is active when phosphorylated.
A phosphatase, which acts constitutively (that is, by default it is always
active) reverses the phosphorylation.
The biological model from~\cite{ferrell,pnasangeliferrellsontag04} is
in~\Fig{mapk_cascades_cell_signaling075_cropped}(b), 
were we wrote
$z_i(t), i=1$$,$$2,$$3$ for MAPK, MAPK-P, and MAPK-PP concentrations
and similarly for the other variables.
The input represents an external signal to this subsystem (typically,
the concentration of a kinase driving forward the reaction).

We make here the simplest assumptions about the dynamics,
amounting basically to a quasi-steady state appproximation of enzyme kinetics.
(For related results using more realistic, mass-action, models,
see~\cite{translation-invariance,06cdc_chemical,persistencePetri}.)
For example, take the reaction shown in the square in 
\Fig{mapk_cascades_cell_signaling075_cropped}(a).
As $y_3$ (MAPKK-PP) facilitates
the conversion of $z_1$ into $z_2$ (MAPK to MAPK-P), the rate of change
$dz_2/dt$ should include a term 
$\alpha (z_1,y_3)$
(and $dz_1/dt$ has a term $-\alpha (z_1,y_3)$)
for some (otherwise unknown) function $\alpha $ 
such that $\alpha (0,y_3)=0$ and
$\frac{\partial \alpha }{\partial z_1}>0$,
$\frac{\partial \alpha }{\partial y_3}>0$ when $z_1>0$.
(Nothing happens if there is no substrate, but more enzyme or more substrate
results in a faster reaction.)
There will also be a term $+\beta (z_2)$ to reflect the phosphatase action.
Similarly for the other species.
The system as given would be represented by a set of seven ordinary
differential equations (or reaction-diffusion PDE's, if spatial localization
is of interest, or delay-differential equations, if appropriate).  

\emph{This system is not monotone} (at least with respect to any orthant
cone), as is easy to verify graphically.  However, as with many other examples
of biochemical networks, \emph{the system is ``monotone in disguise''}, so to
speak, in the sense that a judicious change of variables allows one to apply
MIOS tools.
(Far more subtle forms of this argument are key to applications to
signaling cascades.
A substantial research effort, not reviewed here because of lack of space,
addresses the search for graph-theoretic conditions
that allow one to find such ``monotone systems in disguise''; see
\cite{sysbio04,ejc05es,06cdc_chemical} for references.)

\begin{figure}[h,t]
 \begin{center}
\setlength{\unitlength}{1500sp}%
\begin{picture}(2418,5123)(3593,-3977)
\put(3901,-1261){\circle{600}}
\put(4801,839){\circle{600}}
\put(3901,-2461){\circle{600}}
\put(5701,-2461){\circle{600}}
\put(5703,-3670){\circle{600}}
\put(4769,-377){\circle{600}}
\put(5671,-1264){\circle{600}}
\put(4801,539){\vector( 0,-1){600}}
\put(5701,-2761){\vector( 0,-1){600}}
\put(5701,-1561){\vector( 0,-1){600}}
\put(4201,-1111){\vector( 1, 0){1200}}
\put(5401,-1411){\vector(-1, 0){1200}}
\put(4201,-2311){\vector( 1, 0){1200}}
\put(5401,-2611){\vector(-1, 0){1200}}
\put(4576,-586){\vector(-1,-1){450}}
\put(5026,-586){\vector( 1,-1){450}}
\put(5476,-1561){\vector(-2,-1){1350}}
\put(4801,689){\vector( 0, 1){300}}
\put(4801,-511){\vector( 0, 1){300}}
\put(5701,-1411){\vector( 0, 1){300}}
\put(5701,-2611){\vector( 0, 1){300}}
\put(5701,-3811){\vector( 0, 1){300}}
\put(3901,-1111){\vector( 0,-1){300}}
\put(3901,-2311){\vector( 0,-1){300}}
\put(4876,239){$+$}
\put(5776,-3061){$+$}
\put(5251,-736){$+$}
\put(5776,-1861){$+$}
\put(4276,-736){$-$}
\put(4726,-1861){$-$}
\put(4726,-1111){$-$}
\put(4726,-2311){$-$}
\put(4711,-2739){$-$}
\put(4711,-1516){$-$}
\end{picture}
\caption{Consistent assignment for simple MAPK cascade model}
 \label{mapk-spins}
 \end{center}
 \end{figure}
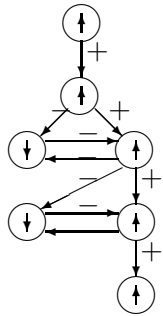

In this example, which in fact was the one whose study initially
led to the definition
of MIOS, the following conservation laws:
$y_1(t)+y_2(t)+y_3(t)\equiv \ytot$ (total MAPKK) and
$z_1(t)+z_2(t)+z_3(t)\equiv \ztot$ (total MAPK)
hold true, assuming no protein turn-over.
This assumption is standard in most of the literature, because transcription
and degradation occur at time scales much slower than signaling.
(There is very recent experimental data that suggests that turn-over
might be fast for some yeast MAPK species.  Adding turn-over would lead to a
different mathematical model.) 
These conservation laws allow us to eliminate variables.
The right trick is to eliminate $y_2$ and $z_2$.
Once we do this,
and write $y_2=\ytot-y_1-y_3$
and
$z_2=\ztot-z_1-z_3$, we are left with the variables $x,y_1,y_3,z_1,z_3$.
For instance, the equations for $z_1,z_3$ look like:
\[
\frac{dz_1}{dt} = -\alpha (z_1,y_3) + \beta (\ztot-z_1-z_3)\quad\quad
\frac{dz_3}{dt} = \gamma (\ztot-z_1-z_3,y_3) -\delta (z_3)
\]
for appropriate increasing
functions $\alpha ,\beta ,\gamma ,\delta $.  The equations for the remaining
variables are similar.
The graph, ignoring, as usual, self-loops (diagonal of Jacobian), is shown in
\Fig{mapk_cascades_cell_signaling075_cropped}(b).
This graph has no negative undirected loops, showing that the (reduced) 
\emph{system is monotone}.  A consistent spin assignment (including the top
input node and the bottom output node) is shown in Figure~\ref{mapk-spins}.
It is also true that this system \emph{has a well-defined monostable state
space response (characteristic)};  there is no space to discuss the proof here,
so we refer the reader to the original papers~\cite{monotoneTAC,monotoneLSU}.

Positive and negative feedback loops around MAPK cascades
have been a topic of interest in the biological literature.
For example, see~\cite{cit13,cit15}
for positive feedback and
\cite{kholodenko00,shvartsman} for negative feedback.
Since we know that the system is monotone and has a characteristic,
MIOS theory as described here can indeed be applied to the example.
We study next the effect of a positive feedback $u=\gain y$
obtained by ``feeding back'' into the input a scalar multiple $\gain$ of the
output.  (This is a somewhat unrealistic model of feedback, since
feedbacks act for example by enhancing the activity of a kinase.
We pick it merely for illustration of the techniques.)

\emph{The theorem does not require actual
  equations for its applicability.}  All that is needed is the knowledge that
 we have a MIOS, and a plot of its characteristic (which, in practice, would be
  obtained from interpolated experimental data).
In order to
illustrate the conclusions, on the other hand, it is worth discussing a
  particular set of equations.  We take equations
and parameters from~\cite{pnasangeliferrellsontag04,sysbio04,ejc05es}:
\beqn
\frac{dx}{dt}  &=& -{\frac{\coefV_2\,x}{\coefK_2+x}} +  \coefV_0 \,\inputu + \coefV_1\\
\frac{dy_1}{dt} &=& {\frac{\coefV_6\,(\ytot-y_1-y_3)}{\coefK_6+(\ytot-y_1-y_3)}}-{\frac{\coefk_3\,x\,y_1}{\coefK_3+y_1}}\\
\frac{dy_3}{dt} &=&
{\frac{\coefk_4\,x\,(\ytot-y_1-y_3)}{\coefK_4+(\ytot-y_1-y_3)}}-{\frac{\coefV_5\,y_3}{\coefK_5+y_3}}\\
\frac{dz_1}{dt} &=& {\frac{\coefV_{10}\,(\ztot-z_1-z_3)}{\coefK_{10}+(\ztot-z_1-z_3)}}-{\frac{\coefk_7\,y_3\,z_1}{\coefK_7+z_1}}\\
\frac{dz_3}{dt} &=&  {\frac{\coefk_8\,y_3\,(\ztot-z_1-z_3)}{\coefK_8+(\ztot-z_1-z_3)}}-{\frac{\coefV_9\,z_3}{\coefK_9+z_3}}
\eeqn
with output $z_3$.
Specifically, we will use the following parameters:
$v_0=0.0015$,
$v_1=0.09$,
$v_2=1.2$,
$v_3=0.064$,
$v_4=0.064$,
$v_5=5$,
$v_6=5$,
$v_7=0.06$,
$v_8=0.06$,
$v_9=5$,
$v_{10}=5$,
$\ytot=1200$,
$\ztot=300$,
$k_2=200$,
$k_3=1200$,
$k_4=1200$,
$k_5=1200$,
$k_6=1200$,
$k_7=300$,
$k_8=300$,
$k_9=300$,
$k_{10}=300$.
(The units are: totals in nM (mol/cm$^3$),
$\coefV$'s in nM$\cdot $sec$^{-1}$ and sec$^{-1}$, 
and $\coefK$'s in nM.)

With these choices,
the steady state step response is the sigmoidal curve shown 
in \Fig{mapk_cascades_cell_signaling075_cropped}(c),
where $y$ is the output $z_3$.
We plotted in the same figure the inverse $g^{-1}$ of the characteristic of
the feedback system, in this case just the linear mapping $y=(1/g)u$,
for three typical ``feedback gains'' ($g$$=$$1/0.98,1/2.1,1/6$).

\begin{figure}[h,t]
 \begin{center}
 \pic{0.2}{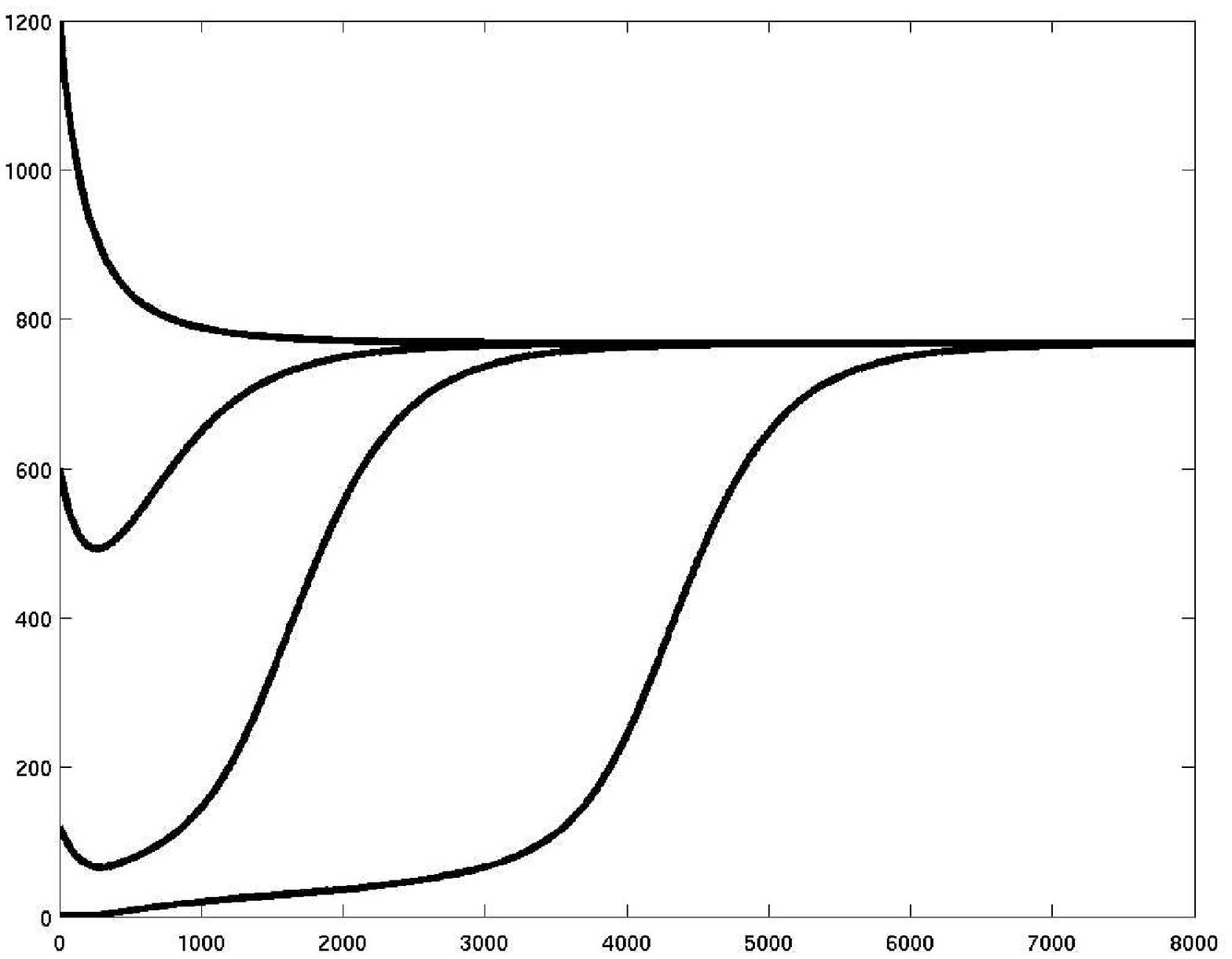}$\!\!\!$
 \pic{0.2}{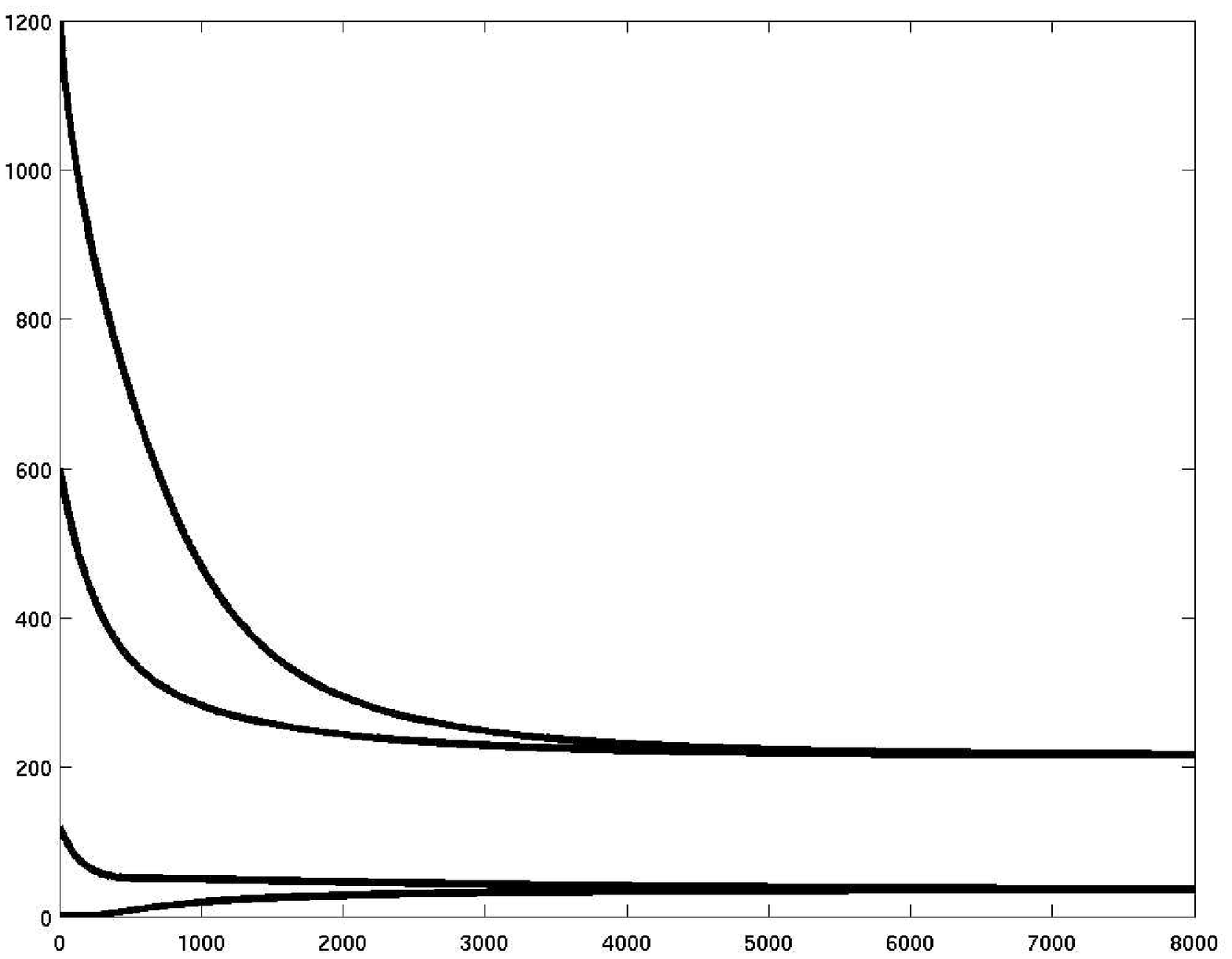}$\!\!\!$
 \pic{0.2}{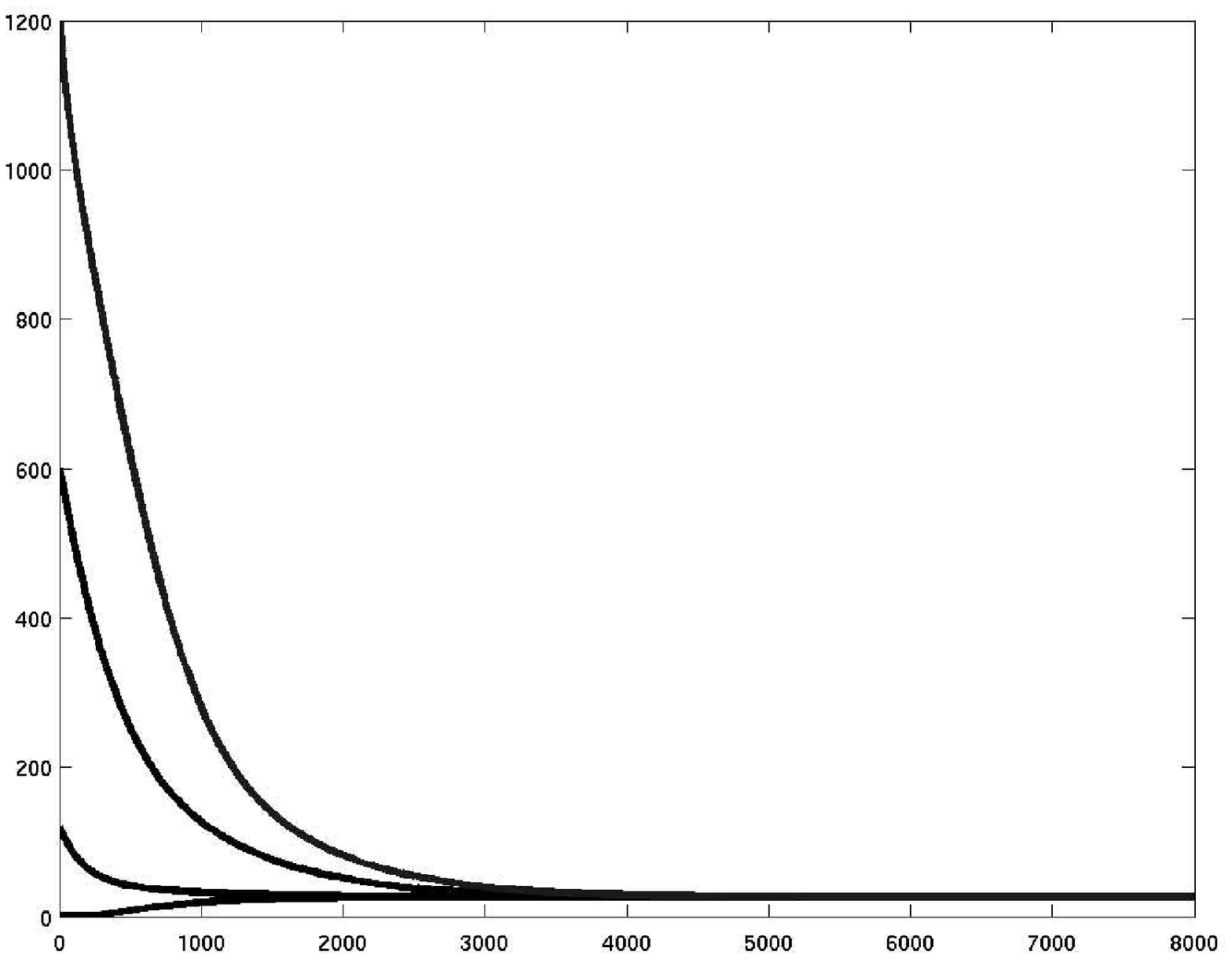}$\!\!\!$
\caption{$\!\!\!\!$(a),(b),(c) $y_3$, $g=1/0.98,1/2.1,1/6$}
 \label{x_coord_high_mapk_5nov04_quality100}
 \end{center}
 \end{figure}

For $g = 1/0.98$ (line of slope 0.98 when plotting $y$ against $u$), there
should be a unique stable state, with a high value of the output $y=z_3$, and
trajectories should generically converge to it. 
Similarly, for $g=1/2.1$ (line of slope 2.1) there should be two stable
states, one with high and one with low $y=z_3$, with trajectories generically
converging to one of these two, because the
line intersects at three points, corresponding to
two stable and one unstable state 
(exactly
as in the discussion concerning the 
simple protein formation/degradation sigmoidal example in
\Fig{hyperbolic_response}(c)).
Finally, for $g=1/6$ (line of slope 6), only the low-$y$ stable state should
persist.
\Fig{x_coord_high_mapk_5nov04_quality100}(a-c) shows plots of the hidden
variable $y_3(t)$ (MAPKK-PP) for several initial states, confirming the
predictions. 
The same convergence results are predicted if there are delays in the feedback
loop, or if concentrations depend on location in a convex spatial domain.
Results for reaction-diffusion PDE's and delay-differential systems are
discussed in~\cite{ejc05es}, and simulation results for this example are also
provided there.

\begin{figure}[h,t]
 \begin{center}
 \pic{0.2}{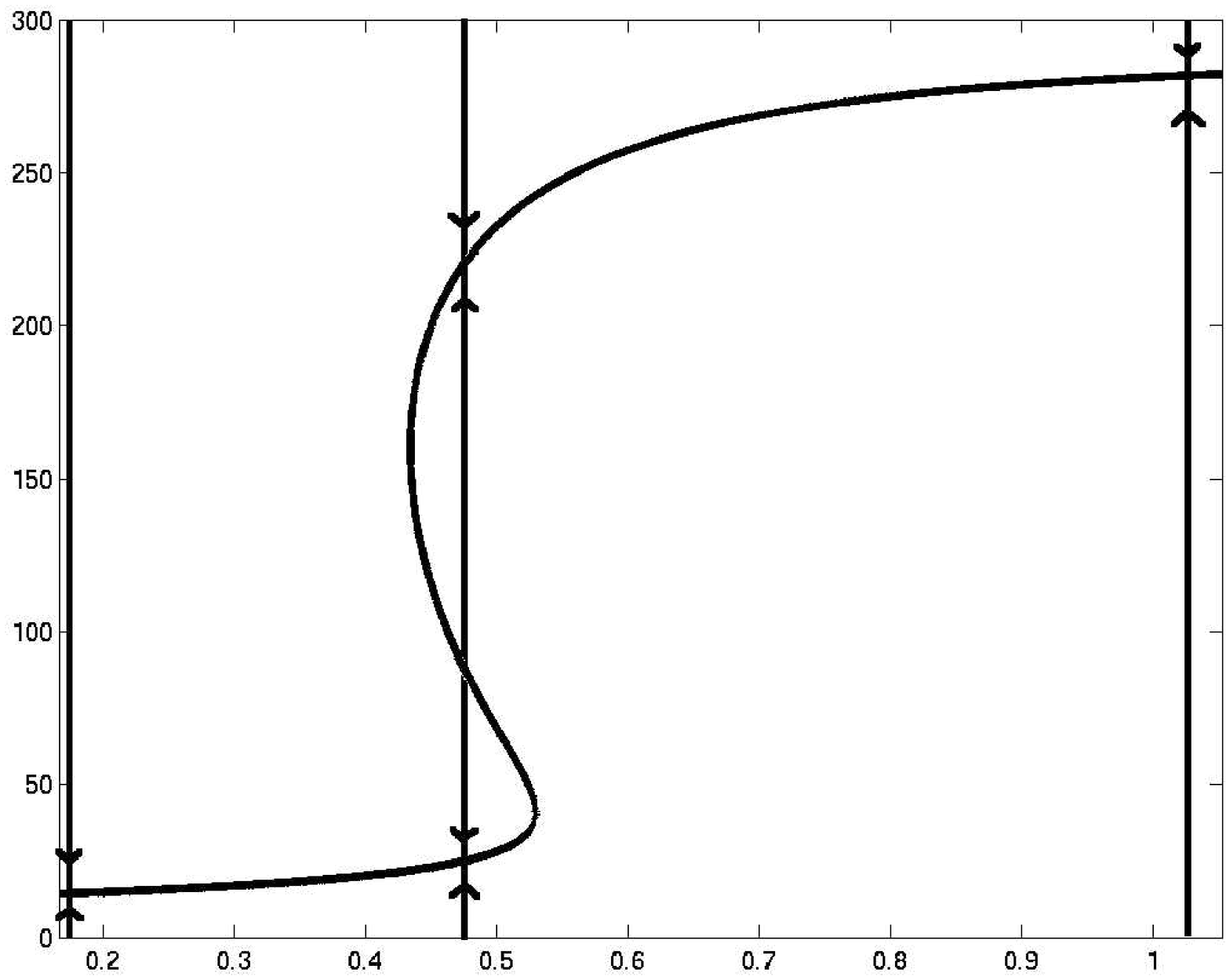}
$\quad$
 \pic{0.2}{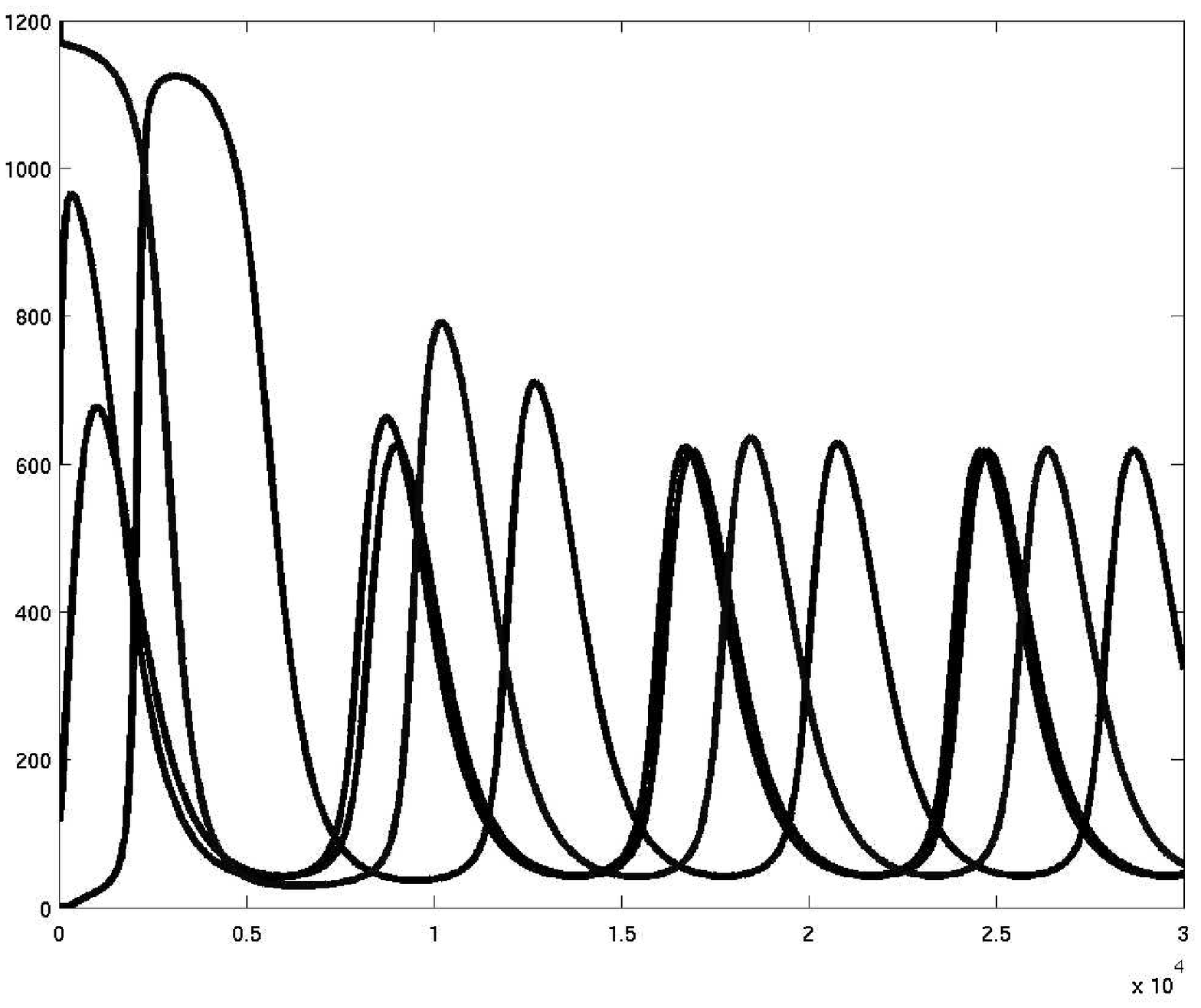}
\caption{(a) bifurcation diagram and relaxation (b) oscillation ($y_3$)}
\label{bif and relax fig}
 \end{center}
 \end{figure}

We may plot the steady state value of $y$, under the feedback $u=gy$,
as the gain $g$ is varied, \Fig{bif and relax fig}(a).
This resulting complete bifurcation diagram showing points of saddle-node
bifurcation can be also completely determined just from the characteristic,
with no need to know the equations of the system.  
Relaxation oscillations may be expected under such circumstances if a
second, slower, feedback loop is used to negatively adapt the gain as a
function of the output.
Reasons of space preclude describing a very general theorem, which shows that
indeed, relaxation oscillations can be guaranteed in this fashion:
see~\cite{gedeon05} for technical details, and~\cite{ejc05es} for a more informal
discussion.
\Fig{bif and relax fig}(b) shows a simulation confirming the
theoretical prediction (details in~\cite{gedeon05,ejc05es}).

\subsubsection*{Negative feedback}

A different set of results apply to inhibitory or negative feedback
interconnections of two MIOS systems (\ref{sys1})-(\ref{sys2}).
A convenient mathematical way to define ``negative feedback'' in the context
of monotone systems is to say that the orders on inputs and outputs are
inverted (example: an inhibition term of the form $\frac{V}{K+y}$, as usual
in biochemistry).  Equivalently, we may incorporate the inhibition into the
output of the second system (\ref{sys2}), which is then seen as an
\emph{anti-monotone} I/O system, and this is how we proceed from now on.
See \Fig{oldfigs78}(c).  We emphasize that the closed-loop systems that
result are \emph{not} monotone, at least with respect to any known order.

The original theorem, from~\cite{monotoneTAC}, is as follows.  We assume once
more that inputs and outputs are scalar ($m$$=$$p$$=$$1$;
see \cite{dcds06} for generalizations).
We once again plot together $k$ and $g^{-1}$, as shown in 
\Fig{oldfigs78}(d).  Consider the following discrete iteration:
\[
u_{i+1}=(g\circ k)(u_i).
\]
Then, if solutions of the closed-loop system are bounded and if
this iteration has a globally attractive fixed point 
${\bar  u}$, as shown in \Fig{oldfigs78}(d),
then the feedback system has a globally attracting steady state.  
(An equivalent condition, see~\cite{dcds06}, is that 
the iteration have no nontrivial period-two orbits.)
We call this result a \emph{small gain theorem} (``SGT''), because
of its analogy to concepts in control theory.

It is easy to see that \emph{arbitrary delays} may be allowed in the feedback
loop. 
In other words, the feedback could be of the form $u(t) = y(t-h)$,
and such delays (even with $h=h(t)$ time varying or even state-dependent, as long
as $t-h(t)\rightarrow \infty $ as $t\rightarrow \infty $) 
\emph{do not destroy global stability of the closed loop.}
In~\cite{enciso_smith_sontagJDE06},
we have now shown also that \emph{diffusion} does not destroy
global stability either.  In other words, a reaction-diffusion system
(Neumann boundary conditions) whose reaction can be modeled in the shown
feedback configuration, has the property that all solutions converge to
a (unique) uniform in space solution.  This is not immediately obvious, since
standard parabolic comparison theorems do not immediately apply to the feedback
system, which is not monotone.

\noindent{\underline{\bf Example: MAPK cascade with negative feedback.}}

\begin{figure}[h,t]
 \begin{center}
 \pic{0.15}{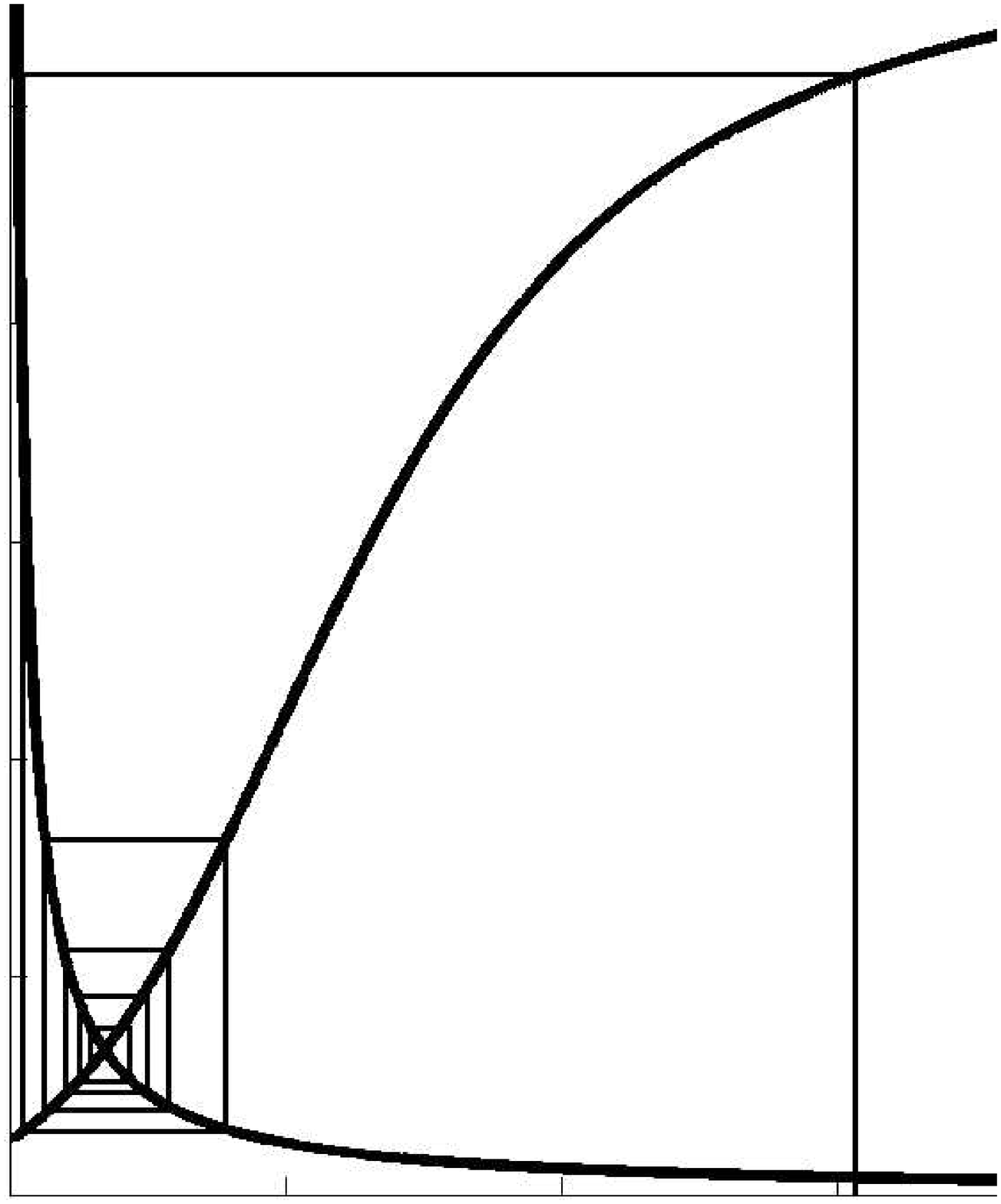}
$\quad$
 \pic{0.18}{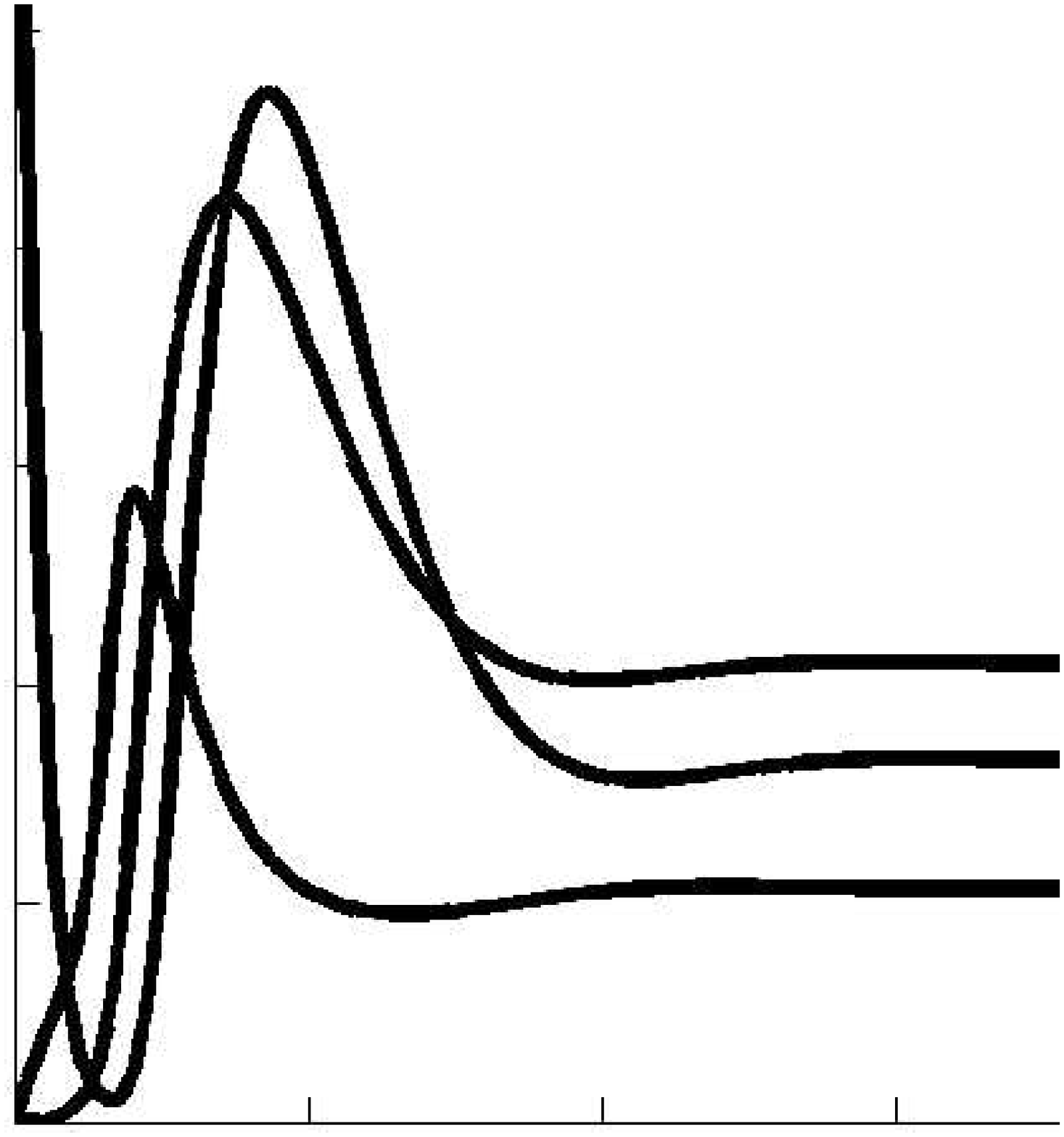}
 \caption{inhibition:~(a) spiderweb and (b) simulation}
\label{inhibition and bifurcation fig}
 \end{center}
 \end{figure}

As with the positive feedback theorem, an important feature is applicability
to highly uncertain systems.  As long as the
component systems are known to be MIOS, the knowledge of I/O response
curves and a planar analysis are sufficient to conclude GAS of the entire
system, which may have an arbitrarily high dimension.  For example, suppose
we take a feedback like $u $$=$$ a$$+$$b/(c$$+$$z_3)$, with a
graph as shown in \Fig{inhibition and bifurcation fig}(a), which
also shows the characteristic and a convergent discrete 1-d iteration
\cite{ejc05es}.
Then, we are guaranteed that all solutions of the closed-loop system converge
to a unique steady state, as confirmed by the simulations
in \Fig{inhibition and bifurcation fig}(b), which shows the
concentrations of the active forms of the kinases.

\noindent{\underline{\bf Example: testosterone model.}}

This example is intended to show that even for a classical mathematical
biology model, a very simple application of the result in~\cite{monotoneTAC}
gives an interesting conclusion.
The concentration of testosterone in the blood of a healthy human male is
known to oscillate periodically with a period of a few hours, in response to
similar oscillations in the concentrations of the luteinising hormone (LH)
secreted by the pituitary gland, and the luteinising hormone releasing hormone
(LHRH), normally secreted by the hypothalamus 
(see~\cite{Cart1986}).  
The well-known textbook~\cite{murray2002} (and its previous editions)
presents this process as an example of a biological oscillator, and proposes a
model to describe it, introducing delays in order to obtain oscillations.
(Since the textbook was written, the physiological mechanism has been much
further elucidated, and this simple model is now known not to be correct.
However, we want merely to illustrate a point about mathematical analysis.)
The equations are:
\beqn
         \dot {R}&=&\frac{A}{K+T}-b_1 R \\
         \dot {L}&=&g_1 R - b_2 L \\
         \dot {T}&=&g_2 L(t-\tau ) - b_3 T 
\eeqn
($R,L,T=$ concentrations
of hormones luteinising hormone releasing, luteinising, and 
testosterone, $\tau =$ delay; we use ``$\dot x$'' to denote time
derivative).
The system may be seen as the feedback connection of
the MIOS system
\beqn
\dot {R}&=&u-b_1 R \\
\dot {L}&=&g_1 R - b_2 L \\
\dot {T}&=&g_2 L - b_3 T 
\eeqn
with the inhibitory feedback $u(t)=g(T-\tau )={A}/({K+T(t-\tau )})$
after moving the delay to the loop (without loss of generality).
The characteristic is linear, $T=k(u) = \frac{g_1g_2}{b_1b_2b_3}u$,
so $g\circ k$ is a fractional transformation $S(u)=\frac{p}{q+u}$.
Since such a transformation has no period-two cycles, global
stability follows.  (For arbitrary, even time-varying, delays.)
This contradicts the existence of oscillations claimed in~\cite{murray2002}
for large enough delays.  (See~\cite{testosterone}, which also explains
the error in~\cite{murray2002}.)

\noindent{\underline{\bf Example: Lac operon.}}
The study of \emph{E.\ Coli} lactose metabolism has been a topic of
research in mathematical biology since Jacob and Monod's classical
work which led to their 1995 Nobel Prize.
For this example, we look at the subsystem modeled in~\cite{mahaffy}.
The lac operon induces production of permease and $\beta $-gal,
permease makes the cell membrane more permeable to lactose,
and genes are activated if lactose present;
lactose is digested by the enzyme $\beta $-gal, and the other
species are degraded at fixed rates.
(In this model from~\cite{mahaffy}, lactose and isolactose are identified, and
catabolic repression by glucose via cAMP is ignored.)
Delays arise from translation of permease and $\beta $-gal.
The equations are:
\beqn
\dot x_1(t)&=&g(x_4(t-\tau )) - b_1x_1(t)
\quad\quad\mbox{lac operon mRNA}
\\
\dot x_2(t)&=&x_1(t)- b_2x_2(t)
\quad\quad\quad\quad\;\;\;\,\mbox{$\beta $-galactoside permease}
\\
\dot x_3(t)&=&rx_1(t) - b_3x_3(t)
\quad\quad\quad\quad\;\;\mbox{$\beta $-galactosidase}
\\
\dot x_4(t)&=& Sx_2(t) - x_3(t)x_4(t)
\quad\quad\quad\;\mbox{lactose}
\eeqn
with $g(x):=(1+Kx^\rho )/(1+x^{\rho })$, $K>1$, and the Hill exponent $\rho $
representing a cooperativity effect.
(All delays have been lumped into one.)
We view this system as a negative feedback loop, where
$u$$=$$x_1$, $v$$=$$x_4$, of a MIOS system (details
in~\cite{dcds06}).
Since there are two inputs and outputs, now we must study the two-dimensional
iteration
\[
(u,v) \;\mapsto \; (g\circ k)(u,v)=
 \left[\frac{g(v)}{b},\frac{Sb_1b_3u}{rb_2g(v)}\right] \,.
\]
Based on results on rational difference equations 
from~\cite{kulenovic}, one concludes that there are no nontrivial 2-periodic
orbits,
provided that
$\rho <(\sqrt{K}$$+$$1)/(K$$-$$1)$, for
arbitrary $b_1,b_2,b_3,r,S$.
Hence, by the theorem, there is a unique steady state of the
original system, which is GAS, even when arbitrary delays are present,

These and other conditions are analyzed in~\cite{dcds06}, where it
is also shown that the results from~\cite{mahaffy} are recovered as
a special case.
Among other advantages of this approach, besides generalizing the result and
giving a conceptually simple proof, we have (because
of~\cite{enciso_smith_sontagJDE06}) the additional conclusion that also for the
corresponding reaction-diffusion system, in which localization is taken
account of, the same globally stable behavior can be guaranteed.

\begin{figure}[h,t]
\minp{0.33}{\pic{0.3}{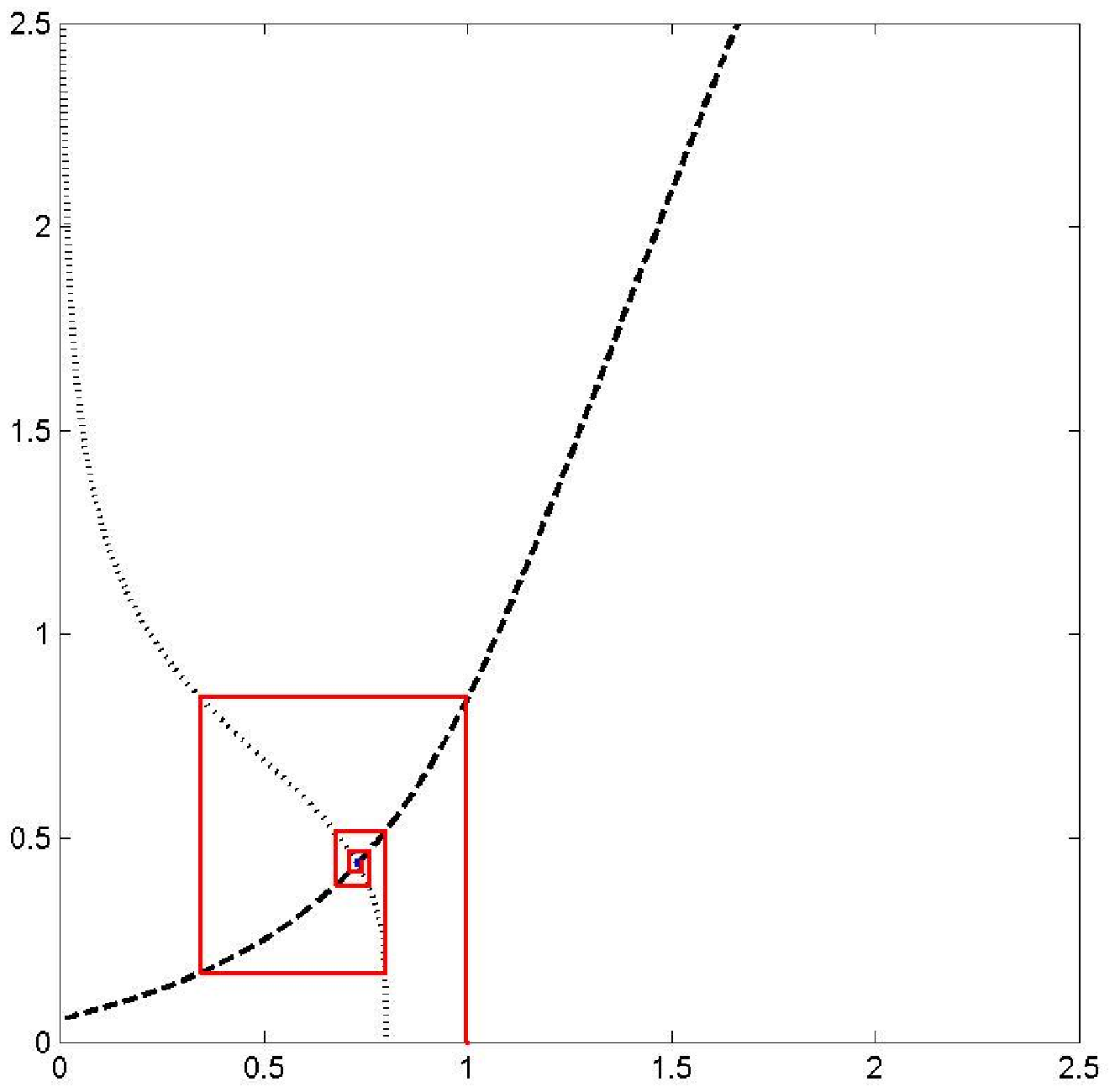}}
\minp{0.33}{\pic{0.3}{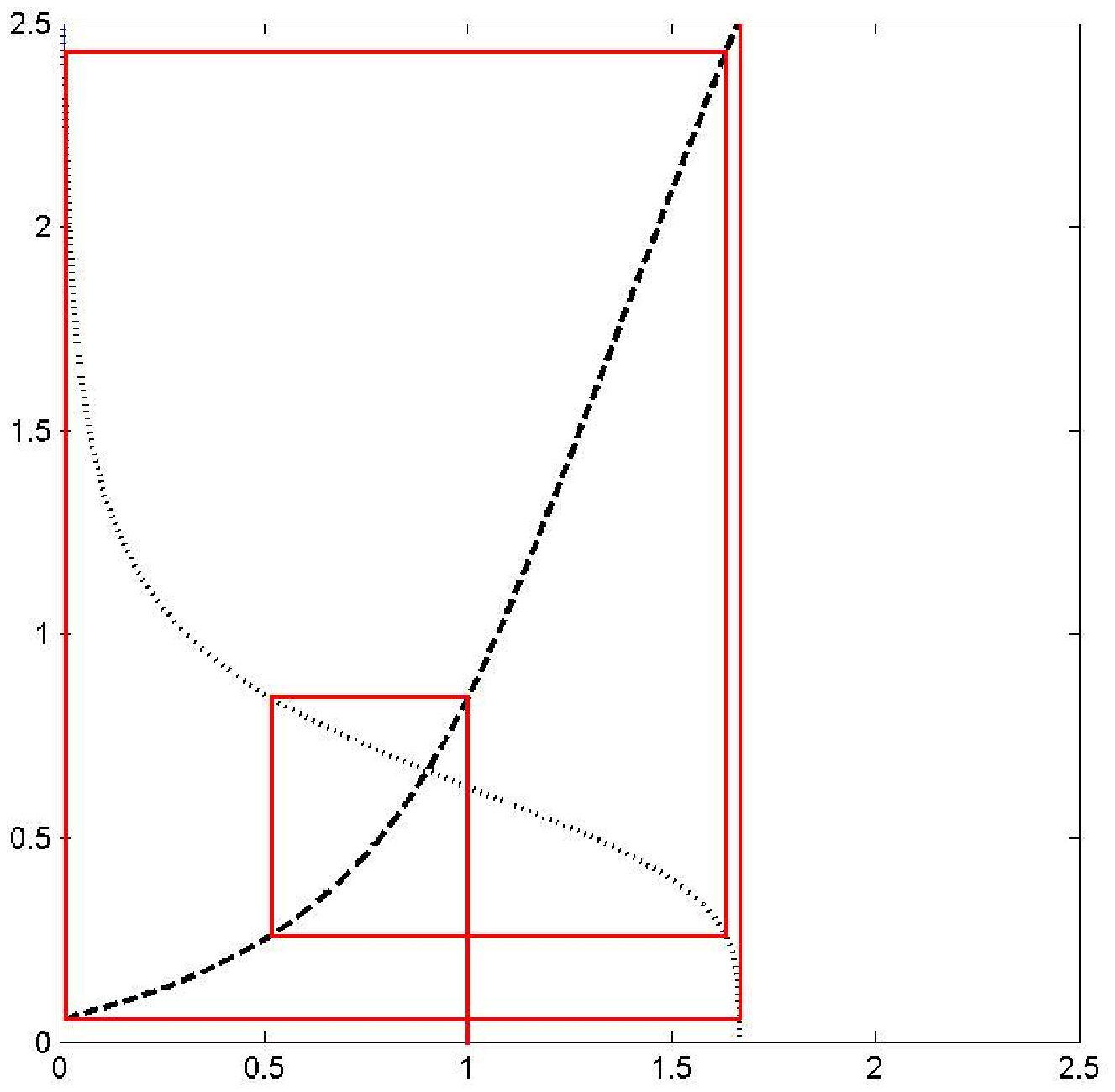}}
\minp{0.33}{\pic{0.3}{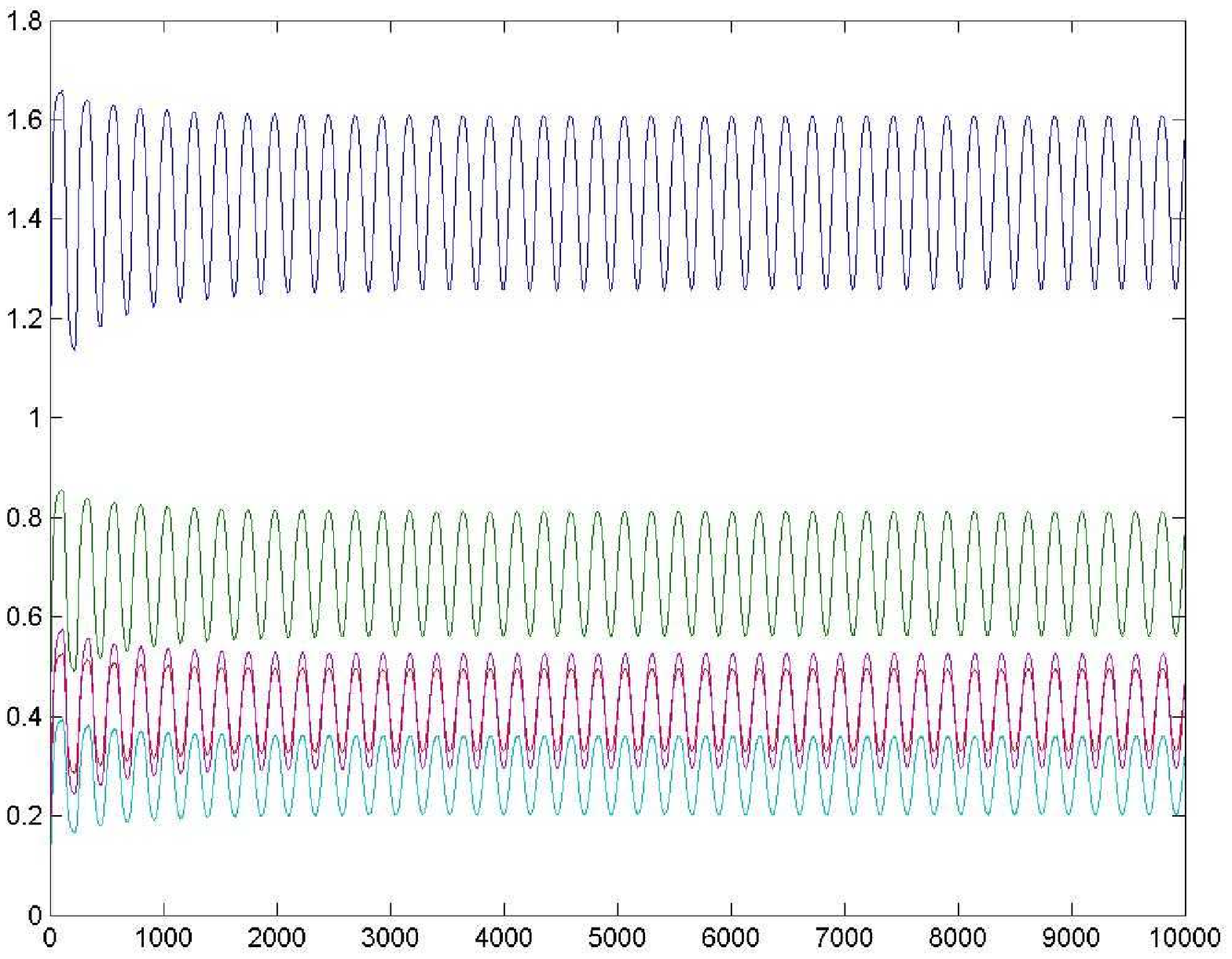}}
 \caption{
 (a) convergent iteration;\hskip1cm
 (b) divergent iteration;\hskip2cm
 (c) oscillations\hskip2cm
}
 \label{circadian-pics}
 \end{figure}

\noindent{\underline{\bf Example: Circadian oscillator.}}
As a final example of the negative feedback theorem, we pick
Goldbeter's
\cite{goldbeter95,goldbeter96} original model of the molecular mechanism
underlying circadian rhythms in \emph{Drosophila}.
(In this oversimplified model, only \emph{per} protein is considered; other
players such as \emph{tim} are ignored.)
PER protein is synthesized at a rate proportional to its mRNA concentration.
Two phosphorylation sites are available, and constitutive
phosphorylation and dephosphorylation occur with saturation dynamics,
at maximum rate $v_i$'s and with Michaelis constants $K_i$.
Doubly phosphorylated PER is degraded, also satisfying
saturation dynamics (with parameters $v_d, k_d$), and it is translocated 
to the nucleus with rate constant $k_1$.
Nuclear PER inhibits transcription of the {\em per\/} gene,
with a Hill-type reaction of cooperativity degree $n$ and threshold constant
$K_I$, and mRNA is produced. and translocated to the cytoplasm, at a rate
determined by a constant $v_s$.  Additionally, there is saturated degradation
of mRNA (constants $v_m$ and $k_m$).
The model is ($P_i$ = \emph{per} phosphorylated at $i$ sites, $P_N$ = nuclear
   \emph{per}, $M$ = {\em per} mRNA):
\beqn
\dot M &=& v_sK_I/(K_I+P_N^n)-v_mM/(k_m+M)\\
\dot P_0 &=& k_sM-V_1P_0/(K_1+P_0)+V_2P_1/(K_2+P_1)\\
\dot P_1 &=& V_1P_0/(K_1+P_0)-V_2P_1/(K_2+P_1)-V_3P_1/(K_3+P_1)+V_4P_2/(K_4+P_2)\\
\dot P_2&=& V_3P_1/(K_3+P_1)-V_4P_2/(K_4+P_2)-k_1P_2+k_2P_N-v_dP_2/(k_d+P_2)\\
\dot P_N&=&k_1P_2-k_2P_N.
\eeqn
Parameters are chosen exactly as in Goldbeter's original paper, except that
the rate $v_s$ of mRNA translocation to the cytoplasm is taken as a 
bifurcation parameter.
The value $v_s=0.76$ from~\cite{goldbeter95} gives oscillatory
behavior.
On the other hand, we may break up the system into the $M$ and $P_i,P_N$
subsystems.
Each of these can be shown to be MIOS and have a characteristic.
(The existence of a characteristic for the $P$-subsystem is nontrivial, and
involves the application 
of Smillie's Theorem~\cite{Sm} for strongly monotone tridiagonal systems, and
more precisely, repeated application of a proof technique in~\cite{Sm}
involving ``eventually monotonicity'' of state variables.)
When $v_s$$=$$0.4$, the discrete iteration is graphically seen to be convergent
(see~\Fig{circadian-pics}(a)),
so the theorem guarantees global
asymptotic stability even when arbitrary delays are introduced in the
feedback.
Bifurcation analysis on delay length and $v_s$ indicates
that local stability will fail for somewhat larger values.
Using again the graphical test, we observe that for
$v_s$$=$$0.5$ there appears 
 limit cycle for the \emph{discrete}
iteration on characteristics, see~\Fig{circadian-pics}(b).
This suggests that oscillations may exist in the full nonlinear differential
equation, at least for appropriate delays lengths.
Indeed, the simulation in~\Fig{circadian-pics}(c) displays such oscillations
(see~\cite{04cdc-circadian,monotoneLSU}).

\subsubsection*{Multivalued Characteristics}

For simplicity, we have not discussed the case when characteristics
are set-valued instead of single-valued.  This general case can also be
productively studied with the toolkit afforded by MIOS interconnection theory,
see~\cite{enciso_multi_submitted,05cdc_enciso_sontag} for positive feedback
and~\cite{leenheer-malisoff} for negative feedback.

\section{Conclusions}

There is a clear need in systems biology to study robust structures and to
develop robust analysis tools.  The theory of monotone systems provides one
such tool.
Interesting and nontrivial conclusions can be drawn from (signed) network
structure alone, which is associated to purely stoichiometric
information about the system, and ignores fluxes.

Associating a graph to a given system,
we may define
spin assignments and consistency,
a notion that may be interpreted also as
non-frustration of Ising spin-glass models.
Every species in a monotone system
(one whose graph is consistent)
responds with a consistent sign to perturbations at every other species.
This property would appear to be desirable in biological networks,
and, indeed, there is some evidence suggesting the near-monotonicity of 
some natural networks.
Moreover, ``near''-monotone systems might be ``practically'' monotone,
in the sense of being monotone under disjoint environmental conditions.

Dynamical behavior of monotone systems is ordered and ``non chaotic''.
Systems close to monotone may be decomposed into a small number of
monotone subsystems,
and such decompositions may be usefully employed to study non-monotone
dynamics as well as to help detect bifurcations even in monotone systems,
based only upon sparse numerical data, resulting in a sometimes useful
model-reduction approach.

\subsubsection*{Acknowledgements}

Much of the author's work on I/O monotone systems was done in collaboration
with David Angeli, as well as Patrick de Leenheer, German Enciso, Bhaskar
Dasgupta, and Hal Smith.  The author also wishes to thank Moe Hirsch, Reka
Albert, Tom Knight, Avi Maayan, Alex van Oudenaarden, and many others, for
useful comments and suggestions regarding the material discussed here.

\end{document}